\documentclass[10pt, conference, letterpaper]{IEEEtran}
\usepackage{cite}
\usepackage{amsmath,amssymb,amsfonts}
\usepackage{algorithm}
\usepackage{algpseudocode}
\usepackage{graphicx}
\usepackage{textcomp}
\usepackage{xcolor}
\usepackage{enumitem}
\usepackage{tikz}
\usepackage{url}
\usetikzlibrary{arrows.meta}
\usepackage{adjustbox}
\usepackage[acronym,toc]{glossaries}
\usepackage{color,soul}
\usepackage{subcaption}
\soulregister\gls{7}
\soulregister\glspl{7}
\usepackage[acronym,nonumberlist]{glossaries-extra}
\makeatletter
\IEEEsettextheight{0.8in}{1.03in}
\IEEEsettopmargin{t}{0.8in}
\IEEEquantizetextheight{c}
\IEEEsettopmargin{q}{0sp}
\makeatother
\makeglossaries
\setabbreviationstyle[acronym]{long-short}
\newacronym{2d}{2D}{two-dimensional}
\newacronym{3d}{3D}{three-dimensional}
\newacronym{3gpp}{3GPP}{3rd generation partnership project}
\newacronym{3msk}{3MSK}{3-level minimum-shift-keying}
\newacronym{4g}{4G}{fourth-generation}
\newacronym{5g}{5G}{fifth-generation}
\newacronym{6dof}{6DoF}{six degrees of freedom}
\newacronym{6g}{6G}{sixth-generation}
\newacronym{6tisch}{6TiSCH}{IPv6 over the TSCH mode of IEEE 802.15.4e}
\newacronym{abp}{ABP}{authentication by personalisation}
\newacronym{ac}{AC}{alternating current}
\newacronym{aci}{ACI}{adjacent channel interference}
\newacronym{aclr}{ACLR}{adjacent channel leakage ratio}
\newacronym{acpr}{ACPR}{adjacent channel power ratio}
\newacronym{adc}{ADC}{analog-to-digital converter}
\newacronym{adr}{ADR}{adaptive data rate}
\newacronym{aec}{AEC}{aluminum electrolytic capacitor}
\newacronym{aes}{AES}{Advanced Encryption Standard}
\newacronym{afe}{AFE}{analog front end}
\newacronym{ag}{AG}{array gain}
\newacronym{agc}{AGC}{automatic gain controller}
\newacronym{agps}{A-GPS}{Assisted Global Positioning System} %bestaat al :) % Aha, dan maak ik er wel Assisted gps van!
\newacronym{ai}{AI}{artificial intelligence}
\newacronym{aimd}{AIMD}{Additive-Increase/Multiplicative-Decrease}
\newacronym{aiot}{A-IoT}{Ambient IoT}
\newacronym{alk}{ALK}{Alkaline}
\newacronym{am}{AM}{amplitude modulation}
\newacronym{amam}{AM/AM}{amplitude modulation to amplitude modulation}
\newacronym{ambc}{AmBC}{ambient backscatter communication}
\newacronym{amc}{AMC}{automatic modulation classification}
\newacronym{amp}{AMP}{AMbient Power}
\newacronym{ampm}{AM/PM}{amplitude modulation to phase modulation}
\newacronym{ao}{AO}{access occasion}
\newacronym{aoa}{AOA}{angle-of-arrival}
\newacronym{aod}{AOD}{angle-of-departure}
\newacronym{ap}{AP}{access point}
\newacronym{api}{API}{application program interface}
\newacronym{apt}{APT}{acoustic power transfer}
\newacronym{apu}{APU}{access point unit}
\newacronym{ar}{AR}{augmented reality}
\newacronym{arima}{ARIMA}{Auto-Regressive Integrated Moving Average}
\newacronym{arp}{ARP}{Antenna Reference Point}
\newacronym{asic}{ASIC}{application specific integrated circuit}
\newacronym{ask}{ASK}{amplitude-shift keying}
\newacronym{at}{AT}{ATtention}
\newacronym{auv}{AUV}{autonomous underwater vehicle}
\newacronym{awg}{AWG}{arbitrary waveform generator}
\newacronym{awgn}{AWGN}{additive white Gaussian noise}
\newacronym{baw}{BAW}{bulk acoustic wave}
\newacronym{bb}{BB}{base-band}
\newacronym{bc}{BC}{backscatter communication}
\newacronym{bcjr}{BCJR}{Bahl-Cocke-Jelinek-Raviv}
\newacronym{bd}{BD}{backscatter device}
\newacronym{be}{BE}{Belgium}
\newacronym{ber}{BER}{bit error rate}
\newacronym{bf}{BF}{beamforming}
\newacronym{bga}{BGA}{ball grid array}
\newacronym{bjt}{BJT}{bipolar junction transistor}
\newacronym{bldc}{BLDC}{brushless DC}
\newacronym{ble}{BLE}{Bluetooth low energy}
\newacronym{bler}{BLER}{block error rate}
\newacronym{bod}{BOD}{Brown-Out Detection}
\newacronym{bom}{BOM}{bill of materials}
\newacronym{bpsk}{BPSK}{binary phase-shift keying}
\newacronym{brp}{BRP}{beam refinement process}
\newacronym{bs}{BS}{base station}
\newacronym{btwt}{bTWT}{broadcast TWT}
\newacronym{bu}{BU}{booster unit}
\newacronym{bw}{BW}{bandwidth}
\newacronym{ca}{CA}{carrier emitter}
\newacronym{cad}{CAD}{channel activity detection}
\newacronym{cars}{CARS}{calibration reference signal}
\newacronym{cbm}{CBM}{condition based maintenance}
\newacronym{cbra}{CBRA}{contention-based random access}
\newacronym{cc}{CC}{constant current}
\newacronym{ccdf}{CCDF}{complementary cumulative distribution function}
\newacronym{ccnn}{CCNN}{circular convolutional neural network}
\newacronym{ccs}{CCS}{correlative channel sounder}
\newacronym{cdf}{CDF}{cumulative distribution function}
\newacronym{cdma}{CDMA}{code division-multiple access}
\newacronym{cdrx}{CDRX}{connected mode DRX}
\newacronym{ce}{CE}{coverage enhancement}
\newacronym{ced}{CED}{cumulative energy density}
\newacronym{ceofdm}{CE-OFDM}{constant-envelope OFDM}
\newacronym{cept}{CEPT}{European Conference of Postal and Telecommunications Administrations}
\newacronym{cf}{CF}{cell-free}
\newacronym{cfmmimo}{CF-mMIMO}{cell-free massive MIMO}
\newacronym{cfo}{CFO}{carrier frequency offset}
\newacronym{cfra}{CFRA}{contention-free random access}
\newacronym{cir}{CIR}{channel impulse response}
\newacronym{cla}{CLA}{closed-loop approach}
\newacronym{clk}{CLK}{clock}
\newacronym{cmos}{CMOS}{complementary metal oxide semiconductor}
\newacronym{cn}{CN}{Core network}
\newacronym{cnn}{CNN}{convolutional neural network}
\newacronym{co}{CO}{continuous operation}
\newacronym{cordic}{CORDIC}{coordinate rotation digital computer}
\newacronym{cots}{COTS}{commercial off-the-shelf}
\newacronym{cp}{CP}{cyclic prefix}
\newacronym{cpe}{CPE}{common phase error}
\newacronym{cpfsk}{CPFSK}{continuous phase frequency shift keying}
\newacronym{cpm}{CPM}{continuous phase modulation}
\newacronym{cpo}{CPO}{carrier phase offset}
\newacronym{cpt}{CPT}{capacitive power transfer}
\newacronym{cpw}{CPW}{coplanar waveguide}
\newacronym{cqi}{CQI}{channel quality indicator}
\newacronym{cr}{CR}{coding rate}
\newacronym{crc}{CRC}{cyclic redundancy check}
\newacronym{crlb}{CRLB}{Cram\'er-Rao lower bound}
\newacronym{crs}{CRS}{cell reference signal}
\newacronym{crtwt}{C-RTWT}{coordinated restricted TWT}
\newacronym{cs}{CS}{compressed sensing}
\newacronym{cse}{CSE}{channel state estimation}
\newacronym{csi}{CSI}{channel state information}
\newacronym{csp}{CSP}{contact service point}
\newacronym{css}{CSS}{chirp spread spectrum}
\newacronym{cu}{CU}{central unit}
\newacronym{cv}{CV}{constant voltage}
\newacronym{cw}{CW}{continuous wave}
\newacronym{d2d}{D2D}{device-to-device}
\newacronym{d2r}{D2R}{device-to-reader}
\newacronym{dab}{DAB}{Digital Audio Broadcasting}
\newacronym{dac}{DAC}{digital-to-analog converter}
\newacronym{daq}{DAQ}{data acquisition system}
\newacronym{das}{DAS}{distributed antenna systems}
\newacronym{db}{DB}{duty-cycled barebone}
\newacronym{dbpsk}{DBPSK}{Differential Binary Phase Shift Keying}
\newacronym{dc}{DC}{direct current}
\newacronym{dcc}{DCC}{dynamic cooperation clustering}
\newacronym{ddc}{DDC}{digital down conversion}
\newacronym{de}{DE}{drain efficiency}
\newacronym{dect}{DECT}{Digital Cordless Telecommunications}
\newacronym{dft}{DFT}{discrete Fourier transform}
\newacronym{dftsofdm}{DFT-s-OFDM}{discrete Fourier Transform spread OFDM}
\newacronym{dftsofdmfdss}{DFT-s-OFDM-FDSS}{DFT-s-OFDM with frequency-domain spectral shaping}
\newacronym{dftsofdmfdssse}{DFT-s-OFDM-FDSS-SE}{DFT-s-OFDM with FDSS and spectral extension}
\newacronym{dl}{DL}{downlink}
\newacronym{dlc}{DLC}{Distributed Laser Charging}
\newacronym{dli}{DLI}{direct link interference}
\newacronym{dlt}{DLT}{Distributed Ledger Technology}
\newacronym{dm}{DM}{diffuse multipath}
\newacronym{dma}{DMA}{Direct Memory Access}
\newacronym{dmac}{DMAC}{Direct Memory Access Controller}
\newacronym{dmc}{DMC}{diffuse multipath component}
\newacronym{dmimo}{D-MIMO}{distributed MIMO}
\newacronym{dnn}{DNN}{deep neural network}
\newacronym{doa}{DOA}{direction-of-arrival}
\newacronym{dp}{DP}{duty-cycled persistent}
\newacronym{dpd}{DPD}{digital pre-distortion}
\newacronym{dpdk}{DPDK}{Data Plane Development Kit}
\newacronym{dram}{DRAM}{dynamic random-access memory}
\newacronym{drcs}{$\Delta$RCS}{differential-radar cross section}
\newacronym{drx}{DRX}{Discontinuous Reception Mode}
\newacronym{dsb}{DSB}{double-sideband}
\newacronym{dsl}{DSL}{digital subscriber line}
\newacronym{dsp}{DSP}{digital signal processing}
\newacronym{dss}{DSS}{Dataset Storage Standard}
\newacronym{duc}{DUC}{digital up-converter}
\newacronym{dvb}{DVB}{Digital Video Broadcasting}
\newacronym{de-ewma}{DE-EWMA}{Dynamic Error-Exponentially Weighted Moving Average}
\newacronym{e2e}{E2E}{end-to-end}
\newacronym{easa}{EASA}{European Union Aviation Safety Agency}
\newacronym{ebg}{EBG}{electromagnetic bandgap}
\newacronym{ebw}{EBW}{excess bandwidth}
\newacronym{ec}{EC}{European Commission}
\newacronym{ecc}{ECC}{Electronic Communications Committee}
\newacronym{ecdf}{eCDF}{empirical cumulative distribution function}
\newacronym{ecdlp}{ECDLP}{elliptic curve discrete logarithm problem}
\newacronym{ecsp}{ECSP}{edge computing service point}
\newacronym{edlc}{EDLC}{electrostatic double-layer capacitors}
\newacronym{edrx}{eDRX}{Extended Discontinuous Reception Mode}
\newacronym{ee}{EE}{energy efficiency}
\newacronym{egprs}{EGPRS}{Enhanced Data Rates for GSM Evolution}
\newacronym{eh}{EH}{energy harvesting}
\newacronym{eirp}{EIRP}{equivalent isotropically radiated power}
\newacronym{em}{EM}{electromagnetic}
\newacronym{embb}{eMBB}{enhanced Mobile Broadband}
\newacronym{en}{EN}{energy neutral}
\newacronym{end}{END}{energy-neutral device}
\newacronym{enob}{ENOB}{effective number of bits}
\newacronym{eol}{EoL}{end of life}
\newacronym{ep}{EP}{energy profiler}
\newacronym{epd}{EPD}{electronic paper display}
\newacronym{epu}{EPU}{edge processing unit}
\newacronym{er}{ER}{Energy Receiver}
\newacronym{erc}{ERC}{European Radiocommunications Committee}
\newacronym{erp}{ERP}{effective radiation power}
\newacronym[plural=ESCs,firstplural=electronic speed controllers (ESCs)]{esc}{ESC}{electronic speed control}
\newacronym{esd}{ESD}{electrostatic discharge}
\newacronym{esl}{ESL}{electronic shelf label}
\newacronym{esr}{ESR}{equivalent series resistance}
\newacronym{et}{ET}{Energy Transmitter}
\newacronym{etsi}{ETSI}{European Telecommunications Standards Institute}
\newacronym{evd}{EVD}{eigenvalue decomposition}
\newacronym{evm}{EVM}{Error Vector Magnitude}
\newacronym{ewlb}{eWLB}{Embedded Wafer Level Ball Grid Array}
\newacronym{ewma}{EWMA}{Exponentially Weighted Moving Average}
\newacronym{eipm}{EIPM}{Energy Intelligence Platform Module}
\newacronym{fa}{FA}{federation anchor}
\newacronym{fair}{FAIR}{Findability, Accessibility, Interoperability, and Reuse of digital assets}
\newacronym{fc}{FC}{fusion center}
\newacronym{fcc}{FCC}{Federal Communications Commission}
\newacronym{fcf}{FCF}{frequency correlation function}
\newacronym{fd}{FD}{front-haul distance}
\newacronym{fdd}{FDD}{frequency-division duplexing}
\newacronym{fde}{FDE}{frequency domain equalizer}
\newacronym{fdm}{FDM}{frequency-division multiplexing}
\newacronym{fdma}{FDMA}{frequency division multiple access}
\newacronym{fdss}{FDSS}{frequency-domain spectral shaping}
\newacronym{fem}{FEM}{finite element analysis}
\newacronym{fembb}{feMBB}{further enhanced mobile broadband}
\newacronym{fft}{FFT}{fast Fourier transform}
\newacronym{fh}{FH}{fronthaul}
\newacronym{fhss}{FHSS}{frequency hopping spread spectrum}
\newacronym{fifo}{FIFO}{First In, First Out}
\newacronym{fim}{FIM}{Fisher information matrix}
\newacronym{fits}{FITS}{Flexible Image Transport System}
\newacronym{fl}{FL}{Federated learning}
\newacronym{fom}{FoM}{figure of merit}
\newacronym{fov}{FOV}{field of view}
\newacronym{fpga}{FPGA}{field-programmable gate array}
\newacronym{fqt}{FQT}{Fully Quantized Training}
\newacronym{fr1}{FR1}{frequency range 1}
\newacronym{fr2}{FR2}{frequency range 2}
\newacronym{fram}{FRAM}{Ferroelectric Random Access Memory}
\newacronym{fsk}{FSK}{frequency shift keying}
\newacronym{fspl}{FSPL}{free space path loss}
\newacronym{fss}{FSS}{frequency selective surface}
\newacronym{gan}{GaN}{gallium nitride}
\newacronym{gb}{GB}{grant-based}
\newacronym{gdpr}{GDPR}{general data protection regulation}
\newacronym{gf}{GF}{grant-free}
\newacronym{gmsk}{GMSK}{Gaussian minimum-shift keying}
\newacronym{gnb}{gNB}{Next Generation Node B}
\newacronym{gni}{GNI}{gross national income}
\newacronym{gnn}{GNN}{graph neural network}
\newacronym{gnss}{GNSS}{global navigation satellite system}
\newacronym{gpclk}{GPCLK}{general purpose clock}
\newacronym{gpio}{GPIO}{General-Purpose Input/Output}
\newacronym{gpl}{GPL}{GNU General Public License}
\newacronym{gprs}{GPRS}{General Packet Radio Services}
\newacronym{gps}{GPS}{Global Positioning System}
\newacronym{gpu}{GPU}{graphical processing unit}
\newacronym{grc}{GRC}{GNU Radio Companion}
\newacronym{gscm}{GSCM}{geometry‐based stochastic model}
\newacronym{gsm}{GSM}{Global System for Mobile Communications}  %
\newacronym{gspm}{GSpM}{Generalized spatial modulation}
\newacronym{gwp}{GWP}{Global Warming Potential}
\newacronym{har}{HAR}{human activity recognition}
\newacronym{harq}{HARQ}{hybrid automatic repeat request}
\newacronym{hat}{HAT}{hardware attached on top}
\newacronym{hcs}{HCS}{human-centric services}
\newacronym{hdf5}{HDF5}{Hierarchical Data Format version 5}
\newacronym{hdr}{HDR}{high data rate}
\newacronym{hfss}{HFSS}{High Frequency Simulator Software}
\newacronym{hf}{HF}{high frequency}
\newacronym{hmd}{HMD}{head-mounted display}
\newacronym{hpbm}{HPBM}{half power beam width}
\newacronym{HyMPRo}{HyMPRo}{Hybrid Multi-Path Routing algorithm}
\newacronym{i.i.d.}{i.i.d.}{independent and identically distributed}
\newacronym{i2c}{I2C}{Inter-Integrated Circuit}
\newacronym{iaq}{IAQ}{Indoor Air Quality}
\newacronym{ib}{IB}{in-band}
\newacronym{ibbc}{IBBC}{inter-band beam configuration}
\newacronym{ibo}{IBO}{input back-off}
\newacronym{ic}{IC}{integrated circuit}
\newacronym{iccs}{ICCS}{Ilmsens correlative channel sounder}
\newacronym{ici}{ICI}{intercarrier interference}
\newacronym{icnirp}{ICNIRP}{International Commission on Non-Ionizing Radiation Protection}
\newacronym{id}{ID}{information decoding}
\newacronym{idft}{IDFT}{inverse discrete Fourier transform}
\newacronym{idxm}{IDXM}{index modulation}
\newacronym{ieee}{IEEE}{Institute of Electrical and Electronics Engineers}
\newacronym{if}{IF}{intermediate-frequency}
\newacronym{ifft}{IFFT}{inverse fast-Fourier-transform}
\newacronym{iid}{i.i.d.}{independently and identically distributed}
\newacronym{iis}{IIS}{integrated information system}
\newacronym{im}{IM}{intermodulation}
\newacronym{imd}{IMD}{intermodulation distortion}
\newacronym{imu}{IMU}{inertial measurement unit}
\newacronym{inh}{InH}{indoor hotspot office}
\newacronym{io}{IO}{input/output}
\newacronym{ioe}{IoE}{Internet of Everything}
\newacronym{iot}{IoT}{Internet of Things}
\newacronym{ipt}{IPT}{inductive power transfer}
\newacronym{ipy}{IPY}{Interventions per Year}
\newacronym{iq}{IQ}{in-phase and quadrature}
\newacronym{iqi}{IQI}{IQ imbalance}
\newacronym{ir}{IR}{infrared}
\newacronym{isac}{ISAC}{integrated sensing and communication} 
\newacronym{isi}{ISI}{intersymbol interference}
\newacronym{ism}{ISM}{industrial, scientific and medical}
\newacronym{isp}{ISP}{internet service provider}
\newacronym{itu}{ITU}{International Telecommunication Union}
\newacronym{itu-r}{ITU-R}{ITU – Radiocommunication Sector}
\newacronym{jesd}{JESD}{Joint Electron Devices Engineering Council}
\newacronym{jfet}{JFET}{junction field effect transistor}
\newacronym{kpi}{KPI}{key performance indicator}
\newacronym{ktofdm}{KT-DFT-s-OFDM}{known-tail-DFT-s-OFDM}
\newacronym{kvi}{KVI}{key value indicator}
\newacronym{larva}{LARVA}{LARge Virtual Array}
\newacronym{lbt}{LBT}{Listen-Before-Talk}
\newacronym{lca}{LCA}{life cycle assessment}
\newacronym{lch}{LCH}{logical channel}
\newacronym{lco}{LCO}{lithium cobalt oxide}
\newacronym{ldo}{LDO}{low-dropout voltage regulator}
\newacronym{ldpc}{LDPC}{low-density parity-check}
\newacronym{ldr}{LDR}{low data rate}
\newacronym{led}{LED}{Light Emitting Diode}
\newacronym{less}{LESS}{Low Energy Scheduler Solution}
\newacronym{lev}{LEV}{light electric vehicle}
\newacronym{lfp}{LFP}{lithium iron phosphate}
\newacronym{lib}{LIB}{Lithium-Ion Battery}
\newacronym{lic}{LIC}{lithium-ion capacitor}
\newacronym{lid}{LID}{Lithium Iron Disulfide}
\newacronym{lidar}{LiDAR}{light detection and ranging}
\newacronym{liion}{Li-ion}{lithium-ion}
\newacronym{lipo}{LiPo}{lithium polymer}
\newacronym{lis}{LIS}{large intelligent surface}
\newacronym{llh}{LLH}{log-likelihood}
\newacronym{lls}{LLS}{link-level simulation}
\newacronym{lmd}{LMD}{Lithium Manganese Dioxide}
\newacronym{lmmse}{LMMSE}{least minimum mean square error}
\newacronym{lmo}{LMO}{lithium ion manganese oxide}
\newacronym{lna}{LNA}{low-noise amplifier}
\newacronym{lo}{LO}{local oscillator}
\newacronym{lora}{LoRa}{long range}
\newacronym{lorawan}{LoRaWAN}{long-range wide-area network}
\newacronym{los}{LoS}{line-of-sight}
\newacronym{lp}{LP}{linear programming}
\newacronym{lpf}{LPF}{low-pass filter}
\newacronym{lpt}{LPT}{laser power transfer}
\newacronym{lpwa}{LPWA}{Low Power Wide Area}
\newacronym{lpwan}{LPWAN}{low-power wide-area network}
\newacronym{lpwans}{LPWANs}{Low-Power Wide-Area Networks}
\newacronym{lqi}{LQI}{link quality indicator}
\newacronym{llr}{LLR}{log-likelihood ratio}
\newacronym{lrelu}{LReLU}{leaky rectified linear unit}
\newacronym{lrt}{LRT}{likelihood-ratio test}
\newacronym{ls}{LS}{least squares}
\newacronym{lsa}{LSA}{large synthetic array}
\newacronym{lsf}{LSF}{large-scale fading}
\newacronym{lsfc}{LSFC}{large-scale fading component}
\newacronym{lstm}{LSTM}{Long Short-Term Memory}
\newacronym{ltc}{LTC}{lithium thionyl chloride}
\newacronym{lte}{LTE}{Long Term Evolution}
\newacronym{ltem}{LTE-M}{Long-Term Evolution Machine Type Communication}
\newacronym{lti}{LTI}{linear time-invariant}
\newacronym{lto}{LTO}{lithium titanate}
\newacronym{lusta}{LUSTA 5G }{Logistique mUltimodale Sécuritaire Téléopérée \& Autonome 5G}
\newacronym{m2m}{M2M}{machine to machine}
\newacronym{ma}{MA}{Movable Antennas}
\newacronym{mac}{MAC}{Medium Access Control}
\newacronym{mate}{MATE}{millimeter-wave MIMO testbed}
\newacronym{mc}{MC}{Monte Carlo}
\newacronym{mcl}{MCL}{Maximum Coupling Loss}
\newacronym{mcs}{MCS}{modulation and coding scheme}
\newacronym{mcu}{MCU}{microcontroller unit}
\newacronym{mec}{MEC}{multi-access edge computing}
\newacronym{mems}{MEMS}{micro-electromechanical systems}
\newacronym{mf}{MF}{matched filter}
\newacronym{milp}{MILP}{Mixed Integer Linear Programming}
\newacronym{mimo}{MIMO}{multiple-input multiple-output}
\newacronym{miso}{MISO}{multiple-input single-output}
\newacronym{ml}{ML}{machine learning}
\newacronym{mlp}{MLP}{multilayer perceptron}
\newacronym{mmic}{MMIC}{monolithic microwave integrated circuit}
\newacronym{mmimo}{mMIMO}{massive MIMO}
\newacronym{mmse}{MMSE}{minimum mean square error}
\newacronym{mmtc}{mMTC}{massive machine-type communications}
\newacronym{mmwave}{mmWave}{millimeter wave}
\newacronym{mn}{MN}{matching network}
\newacronym{mosfet}{MOSFET}{metal-oxide semiconductor field effect transistor}
\newacronym{mpc}{MPC}{multipath component}
\newacronym{mpp}{MPP}{magnetic power profile}
\newacronym{mppt}{MPPT}{maximum power point tracking}
\newacronym{mqtt}{MQTT}{Message Queuing Telemetry Transport}
\newacronym{mr}{MR}{maximum ratio}
\newacronym{mrc}{MRC}{maximum ratio combining}
\newacronym{mrc_em}{MRC}{maximum ratio combining}
\newacronym{mrc_EM}{MRC}{Magnetic Resonance Coupling}
\newacronym{mrt}{MRT}{maximum ratio transmission}
\newacronym{mse}{MSE}{mean square error}
\newacronym{msk}{MSK}{Minimum-Shift Keying}
\newacronym{mtc}{MTC}{Machine-Type Communication}
\newacronym{multi-rat}{Multi-RAT}{multiple radio access technology}
\newacronym{multirat}{Multi-RAT}{Multiple Radio Access Technology}
\newacronym{music}{MUSIC}{MUltiple SIgnal Classification}
\newacronym{nas}{NAS}{Neural Architecture Search}
\newacronym{navauwall}{NAVAUWALL}{AUtomated NAVigation in WALLonia}
\newacronym{nb}{NB}{narrowband}
\newacronym{nbiot}{NB-IoT}{narrowband IoT}
\newacronym{nca}{NCA}{nickel cobalt aluminum}
\newacronym{netcdf}{NetCDF}{Network Common Data Form}
\newacronym{nf}{NF}{noise figure}
\newacronym{nfc}{NFC}{near-field communication}
\newacronym{nfv}{NFV}{network function virtualization}
\newacronym{ngmn}{NGMN}{Next Generation Mobile Networks }
\newacronym{ni}{NI}{National Instruments}
\newacronym{nicd}{NiCd}{nikkel cadmium}
\newacronym{nimh}{NiMH}{nikkel metal hydride}
\newacronym{nlos}{NLoS}{non-line-of-sight}
\newacronym{nmc}{NMC}{nickel manganese cobalt}
\newacronym{nmos}{nMOS}{n-channel metal-oxide semiconductor}
\newacronym{nn}{NN}{neural network}
\newacronym{nnls}{NNLS}{non-negative least squares}
\newacronym{noma}{NOMA}{non-orthogonal multiple access}
\newacronym{np}{NP}{Neyman-Pearson}
\newacronym{npbch}{NPBCH}{Narrowband Physical Broadcast Channel}
\newacronym{npss}{NPSS}{Narrow Band Primay Synchronization Signal}
\newacronym{nr}{NR}{New Radio}
\newacronym{nrs}{NRS}{Narrow Band Reference Signal}
\newacronym{nsss}{NSSS}{Narrowband Secondary Synchronization Signal}
\newacronym{ntp}{NTP}{network time protocol}
\newacronym{oai}{OAI}{OpenAirInterface} % no spelling mistake, it is spelled all glued to eachother...
\newacronym{obw}{OBW}{occupied bandwidth}
\newacronym{odi}{O-DI}{On-Device Intelligence}
\newacronym{oef}{OEF}{Organisation Environmental Footprint}
\newacronym{ofa}{OFA}{Once-for-All}
\newacronym{ofdm}{OFDM}{orthogonal frequency-division multiplexing}
\newacronym{ofdma}{OFDMA}{orthogonal frequency-division multiple access}
\newacronym{ofdmim}{OFDM-IM}{OFDM with index modulation}
\newacronym{olos}{OLoS}{obstructed-line-of-sight}
\newacronym{oma}{OMA}{orthogonal multiple access}
\newacronym{oob}{OOB}{out-of-band}
\newacronym{ook}{OOK}{on-off keying}
\newacronym{opbo}{OPBO}{output power backoff}
\newacronym{oran}{O-RAN}{open radio-access network}
\newacronym{os}{OS}{operating system}
\newacronym{ota}{OTA}{over-the-air}
\newacronym{otaa}{OTAA}{over-the-air authentication}
\newacronym{otaml}{OTA-TinyML}{Over-the-air TinyML}
\newacronym{ova}{OVA}{One-Versus-All}
\newacronym{p1}{P1}{Phase 1}
\newacronym{p2}{P2}{Phase 2}
\newacronym{p2p}{P2P}{point-to-point}
\newacronym{pa}{PA}{power amplifier}
\newacronym{pae}{PAE}{power-added efficiency}
\newacronym{pam}{PAM}{pulse amplitude modulation}
\newacronym{pana}{PanA}{Panel A}
\newacronym{panb}{PanB}{Panel B}
\newacronym{papr}{PAPR}{peak-to-average power ratio}
\newacronym{pb}{PB}{power beacon}
\newacronym{pc}{PC}{pilot count}
\newacronym{pcb}{PCB}{printed circuit board}
\newacronym{pcg}{PCG}{power consumption gain}
\newacronym{pcie}{PCIe}{Peripheral Component Interconnect Express}
\newacronym{pcsi}{PCSI}{perfect channel state information}
\newacronym{pd}{PD}{powered device}
\newacronym{pdcch}{PDCCH}{physical downlink control channel}
\newacronym{pdcp}{PDCP}{packet data convergence protocol}
\newacronym{pdf}{PDF}{probability density function}
\newacronym{pdp}{PDP}{power delay profile}
\newacronym{pdsch}{PDSCH}{physical downlink shared channel}
\newacronym{pe}{PE}{processing element}
\newacronym{peb}{PEB}{positioning error bound}
\newacronym{pef}{PEF}{Product Environmental Footprint}
\newacronym{per}{PER}{packet error rate}
\newacronym{pet}{PET}{privacy enhancing technology}
\newacronym{peet}{PET}{Polyethylene terephthalate}
\newacronym{pg}{PG}{path gain}
\newacronym{pgd}{PGD}{proximal gradient descent}
\newacronym{phy}{PHY}{physical layer}
\newacronym{pin}{PIN}{positive-intrinsic-negative}
\newacronym{pki}{PKI}{public key infrastructure}
\newacronym{pl}{PL}{path loss}
\newacronym{pla}{PLA}{physically large array}
\newacronym{pla2}{PLA}{polylactic acid}
\newacronym{pll}{PLL}{phase-locked loop}
\newacronym[plural=PMs,firstplural=person months (PMs)]{pm}{PM}{person month}
\newacronym{pmf}{PMF}{polymer microwave fiber}
\newacronym{pmic}{PMIC}{power management integrated circuit}
\newacronym{pmu}{PMU}{power management unit}
\newacronym{pn}{PN}{pseudo-noise}
\newacronym{po}{PO}{phase offset}
\newacronym{poc}{PoC}{proof of concept}
\newacronym{poe}{PoE}{power-over-Ethernet}
\newacronym{pps}{1PPS}{pulse per second}
\newacronym{pr}{PR}{phase reversal}
\newacronym{prb}{PRB}{Physical Resource Block}
\newacronym{prbs}{PRBs}{Physical Resource Blocks}
\newacronym{prs}{PRS}{Peripheral Reflex System}
\newacronym{ps}{PS}{Processing System}
\newacronym{psd}{PSD}{power spectral density}
\newacronym{pse}{PSE}{power sourcing equipment}
\newacronym{psk}{PSK}{phase shift keying}
\newacronym{psm}{PSM}{power saving mode}
\newacronym{pss}{PSS}{primary synchronisation signal}
\newacronym{ptp}{PTP}{precision-time protocol}
\newacronym{ptrs}{PTRS}{Phase-Tracking Reference Signals}
\newacronym{ptw}{PTW}{paging time window}
\newacronym{pv}{PV}{photovoltaic}
\newacronym{pw}{PW}{planar wavefront}
\newacronym{pwm}{PWM}{pulse width modulation}
\newacronym{qam}{QAM}{quadrature amplitude modulation}
\newacronym{qos}{QoS}{quality-of-service}
\newacronym{qpsk}{QPSK}{quadrature phase-shift keying}
\newacronym{qrrls}{QR-RLS}{QR decomposition based recursive least squares}
\newacronym{quadriga}{QuaDRiGa}{QUAsi Deterministic RadIo channel GenerAtor}
\newacronym{r2d}{R2D}{reader-to-device}
\newacronym{ra}{RA}{Random Access}
\newacronym{ram}{RAM}{random-access memory}
\newacronym{ran}{RAN}{radio access network}
\newacronym{rar}{RAR}{Random Access Response}
\newacronym{rat}{RAT}{radio access technology}
\newacronym{raw}{RAW}{restricted access window}
\newacronym{rb}{Rb}{Rubidium}
\newacronym{rbs}{RBS}{radio base station}
\newacronym{rbw}{RBW}{resolution bandwidth}
\newacronym{rc}{RC}{raised-cosine}
\newacronym{rcs}{RCS}{radar cross section}
\newacronym{rdl}{RDL}{redistribution layer}
\newacronym{re}{RE}{radio element}
\newacronym{red}{RED}{radio equipment directive}
\newacronym{redcap}{RedCap}{reduced capability}
\newacronym{relu}{ReLU}{rectified linear unit}
\newacronym{rf}{RF}{radio frequency}
\newacronym{rfeh}{RF-EH}{radio frequency energy harvesting}
\newacronym{rfic}{RFIC}{radio-frequency integrated circuit}
\newacronym{rfid}{RFID}{radio frequency identification}
\newacronym{rfpt}{RFPT}{radio frequency power transfer}
\newacronym{rfsoc}{RFSoC}{Radio Frequency System-on-Chip}
\newacronym{rfwpt}{RF-WPT}{radio frequency wireless power transfer}
\newacronym{rir}{RIR}{room impulse response}
\newacronym{ris}{RIS}{reflective intelligent surface}%reconfigurable?
\newacronym{rl}{RL}{reinforcement learning}
\newacronym{rlc}{RLC}{Radio Link Control}
\newacronym{rllmtc}{RLLMTC}{reliable low latency machine type communication}
\newacronym{rls}{RLS}{recursive least squares}
\newacronym{rms}{RMS}{root-mean-square}
\newacronym{rmse}{RMSE}{root-mean-square error}
\newacronym{rmt}{RMT}{random matrix theory}
\newacronym{rof}{RoF}{radio-over-fiber}
\newacronym{ros}{ROS}{robot operating system}
\newacronym{rpi}{RPi}{Raspberry Pi}
\newacronym{rpl}{RPL}{Routing Protocol for Low power and Lossy Networks}
\newacronym{rrc}{RRC}{Radio Resource Connection}
\newacronym{rreq}{RREQ}{route request packet}
\newacronym{rsrp}{RSRP}{Reference Signals Received Power}
\newacronym{rsrq}{RSRQ}{Reference Signal Received Quality}
\newacronym{rss}{RSS}{received signal strength}
\newacronym{rssi}{RSSI}{received signal strength indicator}
\newacronym{rtc}{RTC}{real time clock}
\newacronym{rtf}{RTF}{reader talks first}
\newacronym{rtk}{RTK}{real time kinematics}
\newacronym{rts}{RTS}{ray tracing simulator}
\newacronym{rtwt}{rTWT}{restricted TWT}
\newacronym{ru}{RU}{Radio Unit}
\newacronym{ruc}{rUC}{representative Use Cases}
\newacronym{rv}{RV}{random variable}
\newacronym{rw}{RW}{RadioWeaves}
\newacronym{rx}{RX}{receiver}
\newacronym{rzf}{RZF}{regularized zero forcing}
\newacronym{s-parameter}{S-parameter}{scattering parameter}
\newacronym{sa}{SA}{system aspects}
\newacronym{sa5}{SA}{Stand Alone}
\newacronym{sar}{SAR}{specific absorption rate}
\newacronym{sbl}{SBL}{sparse Bayesian learning}
\newacronym{sc}{SC}{single carrier}
\newacronym{scomp}{SC}{Split Computing}
\newacronym{scfde}{SC-FDE}{single-carrier modulation with frequency-domain-equalization}
\newacronym{scfdma}{SCFDMA}{single-carrier frequency division multiple access}
\newacronym{scs}{SCS}{sub-carrier spacing}
\newacronym{sdap}{SDAP}{service data adaptation protocol}
\newacronym{sdg}{SDG}{Sustainable Development Goal}
\newacronym{sdm}{SDM}{sigma-delta modulator}
\newacronym{sdma}{SDMA}{spatial-division multiple access}
\newacronym{sdn}{SDN}{software-defined network}
\newacronym{sdof}{SDoF}{sigma-delta over fiber}
\newacronym{sdr}{SDR}{software-defined radio}
\newacronym{se}{SE}{spectral efficiency}
\newacronym{sei}{SEI}{specific emitter identification}
\newacronym{ser}{SER}{symbol-error rate}
\newacronym{sf}{SF}{spreading factor}
\newacronym{sfn}{SFN}{single frequency network}
\newacronym{sfo}{SFO}{sampling frequency offset}
\newacronym{sfp}{SFP}{small form-factor pluggable}
\newacronym{sha}{SHA}{Secure Hash Algorithm}
\newacronym{SigMF}{SigMF}{Signal Metadata Format}
\newacronym{sdo}{SDO}{Standards Development Organization}
\newacronym{simo}{SIMO}{single-input multiple-output}
\newacronym{sinr}{SINR}{signal-to-interference-plus-noise ratio}
\newacronym{siso}{SISO}{single-input single-output}
\newacronym{slam}{SLAM}{simultaneous localization and mapping}
\newacronym{slc}{SLC}{spatial leakage suppression}
\newacronym{slerp}{SLERP}{spherical linear interpolation}
\newacronym{sma}{SMA}{SubMiniature version A}
\newacronym{smc}{SMC}{specular multipath component}
\newacronym{smps}{SMPS}{switched mode power supply}
\newacronym{smt}{SMT}{Surface Mount Technology}
\newacronym{sndr}{SNDR}{signal-to-noise-and-distortion ratio}
\newacronym{snidr}{SNIDR}{signal-to-noise-and-interference-and-distortion ratio}
\newacronym{snir}{SNIR}{signal-to-interference-plus-noise ratio}
\newacronym{snr}{SNR}{signal-to-noise ratio}
\newacronym{soc}{SoC}{state of charge}
\newacronym{SoC}{SOC}{System on Chip}
\newacronym{sota}{SotA}{state of the art}
\newacronym{sp}{SP}{service point}
\newacronym{spdt}{SPDT}{single pole double throw}
\newacronym{spi}{SPI}{Serial Peripheral Interface}
\newacronym{spst}{SPST}{single pole single throw}
\newacronym{sram}{SRAM}{static random-access memory}
\newacronym{srd}{SRD}{short-range device}
\newacronym{srls}{SRLS}{standard recursive least squares}
\newacronym{srs}{SRS}{Sounding Reference Signal}
\newacronym{ssb}{SSB}{synchronisation signal block}
\newacronym{ssd}{SSD}{solid state drive}
\newacronym{ssq}{SSQ}{simulator sickness questionnaire}
\newacronym{sta}{STA}{station}
\newacronym{steam}{STEAM}{science, technology, engineering, the arts, and mathematics}
\newacronym{svd}{SVD}{singular value decomposition}
\newacronym{sw}{SW}{spherical wavefront}
\newacronym{swipt}{SWIPT}{simultaneous wireless information and power transfer}
\newacronym{synce}{SyncE}{Synchronous Ethernet}
\newacronym{ta}{TA}{timing advance}
\newacronym{tau}{TAU}{tracking area update}
\newacronym{tcer}{TCER}{transported to consumed energy ratio}
\newacronym{tcp}{TCP}{Transmission Control Protocol}
\newacronym{tcxo}{TCXO}{temperature compensated crystal oscillator}
\newacronym{tdce}{TD-CE}{time-domain compression and expansion}
\newacronym{tdd}{TDD}{time division duplexing}
\newacronym{tdma}{TDMA}{time division-multiple access}
\newacronym{tdoa}{TDOA}{time-difference-of-arrival}
\newacronym{teg}{TEG}{Thermoelectric Generator}
\newacronym{tsg}{TSG}{Technical Specification Group}
\newacronym{thz}{THz}{Terahertz}
\newacronym{tinyml}{TinyML}{Tiny Machine Learning}
\newacronym{tinytl}{TinyTL}{Tiny Transfer Learning}
\newacronym{tinyol}{TinyOL}{Tiny Online Learning}
\newacronym{tinyrl}{TinyRL}{Tiny Reinforcement Learning}
\newacronym{tl}{TL}{Transfer Learning}
\newacronym{to}{TO}{timing offset}
\newacronym{toa}{TOA}{time-of-arrival}
\newacronym{tof}{ToF}{time-of-flight}
\newacronym{tosm}{TOSM}{through-open-short-match}
\newacronym{tpms}{TPMS}{Tire-Pressure Monitoring System}
\newacronym{trl}{TRL}{technology readyness level}
\newacronym{trp}{TRP}{Transmission Reception Point}
\newacronym{teng}{TENG}{Triboelectric nanogenerators}
\newacronym{tsn}{TSN}{time-sensitive networking}
\newacronym{ttf}{TTF}{tag talks first}
\newacronym{ttff}{TTFF}{Time To First Fix}
\newacronym{tti}{TTI}{transmission time interval}
\newacronym{ttm}{TTM}{time to market}
\newacronym{ttn}{TTN}{The Things Network}
\newacronym{tx}{TX}{transmitter}
\newacronym{twt}{TWT}{Target Wake Time}
\newacronym{uart}{UART}{Universal Asynchronous Receiver/Transmitter}
\newacronym{uav}{UAV}{unmanned aerial vehicle}
\newacronym{uc}{UC}{use case}
\newacronym{ucie}{UCIe}{Universal Chiplet Interconnect Express}
\newacronym{udp}{UDP}{User Datagram Protocol}
\newacronym{ue}{UE}{user equipment}
\newacronym{ugv}{UGV}{unmanned ground vehicle}
\newacronym{uhd}{UHD}{USRP hardware driver}
\newacronym{uhf}{UHF}{ultra-high frequency}
\newacronym{ul}{UL}{uplink}
\newacronym{ula}{ULA}{uniform linear array}
\newacronym{ule}{ULE}{ultra low energy}
\newacronym{ulp}{ULP}{ultra-low power}
\newacronym{uMIMO}{$\mu$-MIMO}{ultra-massive Multiple-Input Multiple-Output}
\newacronym{ummtc}{umMTC}{ultra massive machine type communication}
\newacronym{un}{UN}{United Nations}
\newacronym{unow}{uNOW}{unified non-orthogonal waveform}
\newacronym{upa}{UPA}{uniform planar array}
\newacronym{up}{UP}{ultra passive}
\newacronym{ura}{URA}{uniform rectangular array}
\newacronym{urllc}{URLLC}{ultra-reliable low-latency communications}
\newacronym{usrp}{USRP}{universal software radio peripheral}
\newacronym{uv}{UV}{unmanned vehicle}
\newacronym{uw}{UW}{unique word}
\newacronym{uwb}{UWB}{ultrawideband}
\newacronym{uwofdm}{UW-DFT-s-OFDM}{unique word DFT-s-OFDM}
\newacronym{v2v}{V2V}{vehicle-to-vehicle}
\newacronym{vco}{VCO}{voltage-controlled oscillator}
\newacronym{vep}{VEP}{virtual edge platform}
\newacronym{vlc}{VLC}{visible light communication}
\newacronym{vlp}{VLP}{visible light positioning}
\newacronym{vna}{VNA}{vector network analyzer}
\newacronym{voc}{VOC}{Voltatile Organic Compound}
\newacronym{votable}{VOTable}{Virtual Observatory Table}
\newacronym{vr}{VR}{virtual reality}
\newacronym{vuca}{VUCA}{volatile, uncertain, complex and ambiguous}
\newacronym{vswr}{VSWR}{Voltage Standing Wave Ratio}
\newacronym{wifi}{Wi-Fi}{Wireless Fidelity}
\newacronym{wirelesshart}{WirelessHART}{Wireless Highway Addressable Remote Transducer Protocol}
\newacronym{wur}{WuR}{wake-up radio}
\newacronym{wus}{WuS}{wake-up signal}
\newacronym{wb}{WB}{wideband}
\newacronym{wban}{WBAN}{wireless body area network}
\newacronym{wd}{WD}{Wireless distance}
\newacronym{wimax}{WiMAX}{Worldwide Interoperability for Microwave Access}
\newacronym{wlan}{WLAN}{wireless LAN}
\newacronym[plural=WPs,firstplural=work packages (WPs)]{wp}{WP}{work package}
\newacronym{wpc}{WPC}{Wireless Power Consortium}
\newacronym{wpt}{WPT}{wireless power transfer}
\newacronym{wr}{WR}{White Rabbit}
\newacronym{wrsn}{WRSN}{wireless rechargeable sensor network}
\newacronym{wsn}{WSN}{Wireless Sensor Network}
\newacronym{wcma}{WCMA}{weather conditioned moving average}
\newacronym{xets}{XETS}{cross exponentially tapered slot}
\newacronym{xlmimo}{XL-MIMO}{extremely large-scale MIMO}
\newacronym{xr}{XR}{extended reality}
\newacronym{z3ro}{Z3RO}{zero third-order distortion}
\newacronym{zed}{ZED}{Zero-Energy device}
\newacronym{zf}{ZF}{zero-forcing}
\newacronym{zmcscg}{ZMCSCG}{zero mean circularly symmetric complex Gaussian}
\newacronym{zmq}{ZMQ}{ZeroMQ}
\newacronym{ztofdm}{ZT-DFT-s-OFDM}{zero-tail DFT-s-OFDM}

\newacronym{mobc}{MoBC}{monostatic backscatter communication}

\newacronym{bibc}{BiBC}{bistatic backscatter communication}
\newacronym{t2t}{T2T}{tag-to-tag}
\newacronym{pls}{PLS}{physical layer security}

\newacronym{cea}{CEA}{controlled environment agriculture}
\newacronym{abd}{ABD}{ambient-backscattering device}
\newacronym{keh}{KEH}{kinetic energy harvester}
\newacronym{svm}{SVM}{support vector machine}
\newacronym{wisp}{WISP}{Wireless Identification and Sensing Platform} 

\begin{document}

\title{Adaptive Contention-based Random Access for Uplink Reporting in 3GPP Ambient IoT Networks}

% author names and affiliations
% use a multiple column layout for up to three different
% affiliations
\author{\IEEEauthorblockN{1\textsuperscript{st} David E. Ru\'{i}z-Guirola}
\IEEEauthorblockA{\textit{CWC-}
\textit{University of Oulu}\\
Oulu, Finland \\
David.RuizGuirola@oulu.fi}
\and
\IEEEauthorblockN{2\textsuperscript{nd} Samer Nasser}
\IEEEauthorblockA{\textit{University of Antwerp-imec} \\
Belgium \\
samer.nasser@uantwerpen.be}
\and
\IEEEauthorblockN{3\textsuperscript{rd} Bikramjit Singh}
\IEEEauthorblockA{\textit{LMF Ericsson} \\
Jorvas, Finland \\
bikramjit.b.singh@ericsson.com}
\and
\IEEEauthorblockN{4\textsuperscript{th} Henrique Duarte Moura}
\IEEEauthorblockA{\textit{University of Antwerp-imec} \\
Belgium \\
henrique.duartemoura@imec.be}
\and
\IEEEauthorblockN{5\textsuperscript{th} Andrey Belogaev}
\IEEEauthorblockA{\textit{University of Antwerp-imec} \\
Belgium \\
andrei.belogaev@uantwerpen.be}
\and
\IEEEauthorblockN{6\textsuperscript{th} Jeroen Famaey}
\IEEEauthorblockA{\textit{University of Antwerp-imec} \\
Belgium \\
jeroen.famaey@imec.be}
\and
\IEEEauthorblockN{7\textsuperscript{th} Efstathios Katranaras}
\IEEEauthorblockA{\textit{Sequans Communications}\\
UK\\
ekatranaras@sequans.com
}
\and
\IEEEauthorblockN{8\textsuperscript{th} Mahdi Shahabi}
\IEEEauthorblockA{\textit{Sequans Communications}\\
UK\\
mshahabi@sequans.com
}
\and
\IEEEauthorblockN{9\textsuperscript{th} Onel L. A. L\'{o}pez}
\IEEEauthorblockA{\textit{CWC-}
\textit{University of Oulu}\\
Oulu, Finland \\
Onel.AlcarazLopez@oulu.fi
}

\thanks{This work has been partially supported by the Research Council of Finland (Grants 369116 (6G Flagship Programme) and 362782 (ECO-LITE)), the Finnish Foundation for Technology Promotion, and the European Commission through the Horizon Europe/JU SNS project AMBIENT-6G (Grant 101192113).}
}

% make the title area
\maketitle

% As a general rule, do not put math, special symbols or citations
% in the a\gls{bs}tract
\begin{abstract}
Ambient Internet of Things (A-IoT) targets energy harvesting (EH), battery-less devices as a simple connectivity solution for extensive ultra-low-power deployments.  
These devices typically face intermittent energy availability, making uplink reports increasingly susceptible to access collisions and energy outages. In this paper, we build upon the cellular standardization of A-IoT and examine the paging-triggered contention-based random access (CBRA) framework for uplink reporting. We analyze the effects of energy availability and collisions on these systems and introduce an EH-aware access control mechanism. In this mechanism, the reader broadcasts an access probability in the paging message, which helps regulate the number of devices attempting random access. Results show that, unlike the baselines, the proposed method scales well under dense deployments by keeping collisions nearly constant, improving access efficiency, and substantially reducing the number of paging rounds required for successful reporting.  
These results highlight the importance of lightweight reader-side access control for reliable and resource-efficient reporting in A-IoT environments. 
\end{abstract}

\IEEEpeerreviewmaketitle

\section{Introduction}
%\IEEEPARstart{T}{here} is a growing demand for ultra-low power and ultra-low complexity devices that require maintenance-free and battery-less operation~\cite{singh2025physical}. 
The next wave of massive \gls{iot} will involve ultra-low-power devices that operate under $\mu$W power budgets and are powered by \gls{eh}. This development has motivated the adoption of \gls{aiot} within the \gls{3gpp} to enable maintenance-free and energy-neutral operation~\cite{singh2025physical}. These devices will be capable of short bursts of signaling and data exchange, while supporting reduced coverage and complexity compared to \gls{nbiot}, and \gls{redcap}~\cite{lopez2024zero,qu2024ambient}. 

In \gls{aiot}, communication starts with the reader sending a paging signal, \textit{i.e.}, a \gls{r2d} phase. After receiving this signal, devices compete for uplink access, \textit{i.e.}, \gls{d2r} phase, using two types of \gls{ra} methods: \gls{cbra} and \gls{cfra}, which depend on predefined \glspl{ao}. 
\gls{cfra} is suitable when paging a single device or a small group of devices, as it allows the reader to pre-assign resources individually, thereby reducing overall latency in the \gls{ra} process. However, as the number of devices increases, pre-assigning resources becomes inefficient. In situations where the reader needs to trigger multiple devices, \gls{cbra} is a more scalable option~\cite{singh2025physical}. The basic \gls{cbra} procedure follows the principles of slotted ALOHA, where devices randomly select \glspl{ao}. If there is a collision, the access attempt fails, requiring the device to try again during subsequent paging opportunities~\cite{shen2025state,van2025higher}.

\begin{figure}[t!]
\centerline{\includegraphics[width=0.9\columnwidth]{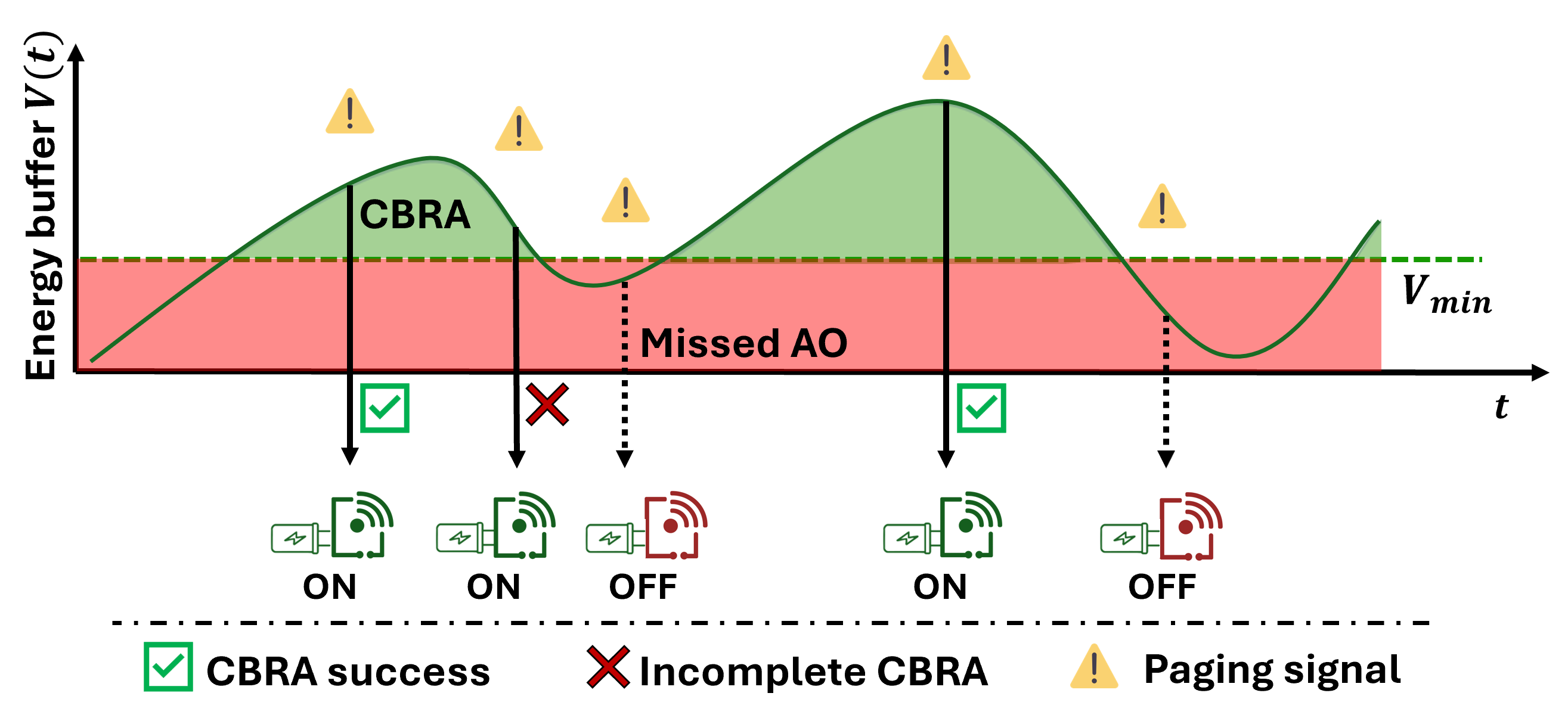}}
%\vspace{2mm}
\caption{Illustration of \gls{eh}-driven \gls{aiot} device availability. The capacitor voltage evolves over time, influencing the device's ability to respond to paging and \glspl{ao}. When $V(t)\geq V_{\min}$, the device is ON and can attempt CBRA.  Conversely, when $V(t)<V_{\min}$, the device remains unavailable, which may lead to missed \glspl{ao}. Even when the device is energy-available, the access attempt may still fail due to collision.}
\vspace{-3mm}
\label{fig:energy_buffer}
\end{figure}

A significant challenge in \gls{aiot} is that the availability of devices is heavily influenced by their energy storage status, which is often duty-cycled. This means that devices may miss paging signals as they strive to harvest enough energy for future \glspl{ao}~\cite{kim2025challenges}. Moreover, even if they do receive a paging signal, they may be unable to complete the \gls{cbra} due to energy outages or collisions with other device transmissions. As a result, devices often make repeated access attempts during times of network congestion, which can further deplete their stored energy. These issues can lead to increased latency and wasted energy, while \gls{eh}-driven device unavailability can severely affect paging reception, access completion, and inventory latency~\cite{kim2025challenges,wu2025fast}.
\figurename~\ref{fig:energy_buffer} illustrates this behavior by mapping the capacitor-voltage evolution to the device availability state and to its ability to react to paging-triggered \glspl{ao}.

Existing studies on \gls{aiot} have focused on \gls{3gpp} standardization aspects, including the paging-triggered \gls{cbra}~\cite{singh2025physical,shen2025state,van2025higher,qu2024ambient}. However, previous research has mainly described these standardized access mechanisms, whereas reader-side control strategies that explicitly account for intermittent device availability to regulate \gls{cbra} contention remain largely unexplored. In this paper, we address this gap by proposing an \gls{eh}-aware access probability control mechanism for \gls{cbra} uplink reporting that remains aligned with the ongoing \gls{3gpp} \gls{aiot} framework, while improving scalability and reporting reliability under dense deployments. Specifically, we model asynchronous paging-triggered \gls{cbra} uplink reporting for \gls{eh}-powered \gls{aiot} devices under Bernoulli reader requests. We also provide a compact approximation for the per-round \gls{cbra} success probability. Additionally, we propose an \gls{eh}-aware access-probability control per \gls{ao}, broadcast via paging, that improves reliability/efficiency compared to a fixed standard \gls{cbra}.
The main findings show that the proposed \gls{eh}-aware access control maintains a nearly constant collision probability as the number of devices increases, in contrast to the rapid increase in collisions of the baseline scheme. Moreover, the proposed method improves resource efficiency and reduces the number of paging rounds required for successful reporting.

The rest of the paper is organized as follows. Section~\ref{sec:system} presents the system model and random access procedure. Section~\ref{sec:ra} outlines \gls{cbra} policies and presents the proposed \gls{eh}-aware access control. The proposal is evaluated through numerical simulations in Section~\ref{sec:results}. Finally, we conclude the paper in Section~\ref{sec:conclusions}. 

\section{System Model}\label{sec:system}

\begin{figure}[t!]
\centerline{\includegraphics[width=0.9\columnwidth]{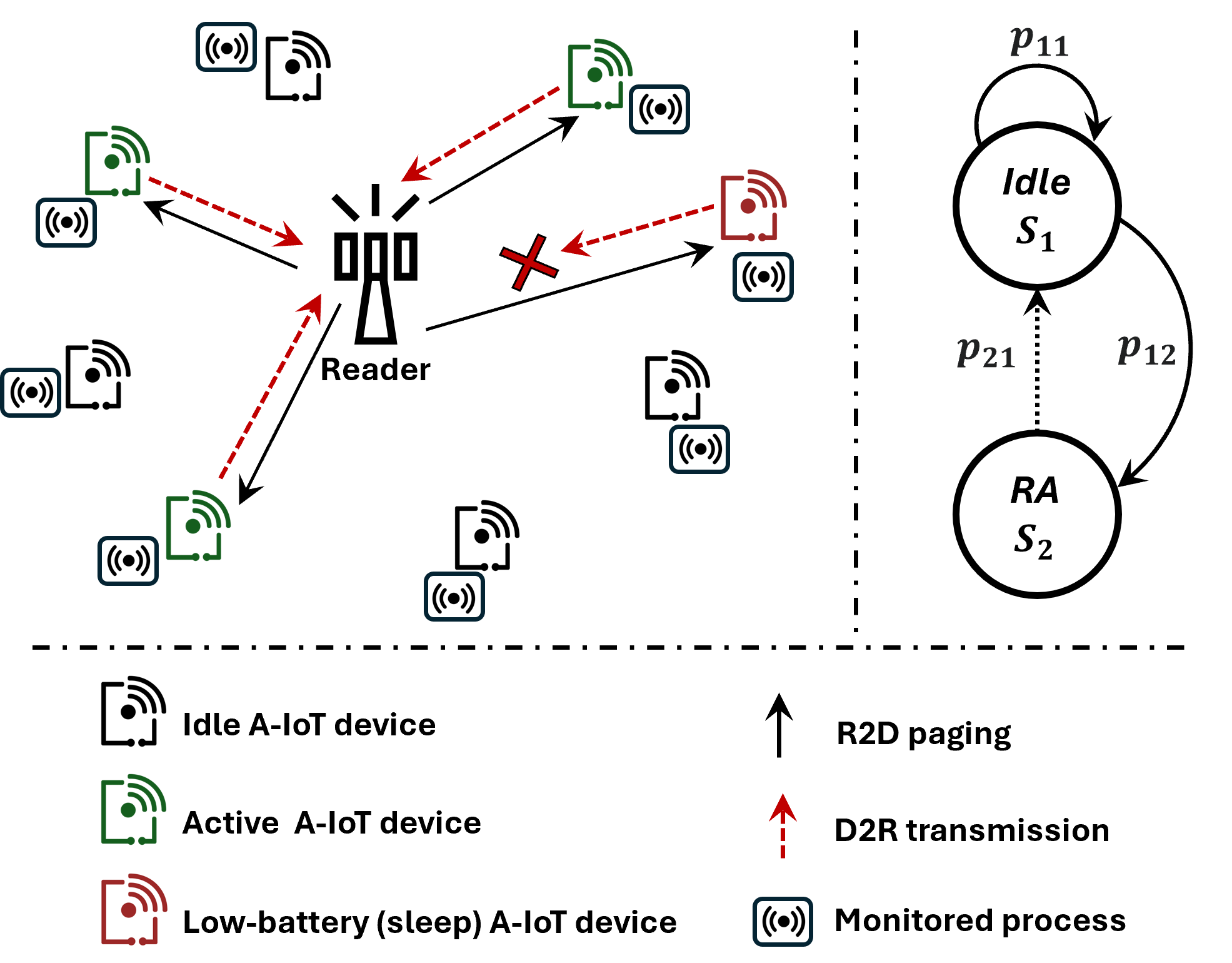}}
%\centerline{\includegraphics[width=\columnwidth]{fig/Energy_buffer.png}}
%\vspace{-2mm}
\caption{Illustration of an \gls{aiot} network where the reader controls the paging for \gls{ra} and collects information from various devices. The \gls{aiot} device's operation states are modeled as a two-state discrete-time Markov chain. Reader requests follow  a Bernoulli process with probability $p_{\mathrm{r2d}}$.}
\vspace{-3mm}
\label{fig:system_model}
\end{figure}

Consider the \gls{aiot} scenario depicted in~\figurename~\ref{fig:system_model}, a set ${\mathcal{N}}$ of $N$ \gls{aiot} devices with \gls{eh} capabilities reports to a gateway/reader. 
The devices are deployed across the coverage area to sense local processes and remain inactive and monitoring for paging until prompted by the network via \gls{wur}. Upon receiving the reader request/polling, the devices respond by transmitting event-related information to the reader, e.g., sensor readings. 

Herein, we assume that time is measured in \glspl{tti} and information request/polling events are generated according to a Bernoulli process with probability $(p_{\mathrm{r2d}})$ per simulation step. This refers to \emph{`when'} the reader sends out a polling message to all connected devices and not to any particular device. When a polling request occurs, devices generate a report and send it to the reader. A device is considered \emph{eligible} to complete the reporting procedure only if its capacitor voltage at the polling instant satisfies $V \geq V_{\min}$, where $V_{\min}$ is the minimum voltage required to sustain sensing and the \gls{d2r} transmission. 

\subsection{\gls{cbra}}\label{subsec:cbra}

\begin{figure*}[t]
\centering
\begin{adjustbox}{width=0.8\linewidth}
\begin{tikzpicture}[
    >=Stealth,
    thick,
    font=\small,
    msg/.style={align=center, font=\small},
    noteR/.style={align=left, anchor=west, font=\footnotesize, text width=5.4cm},
    noteL/.style={align=right, anchor=east, font=\footnotesize, text width=5.0cm},
]
  % Nodes
  \node (r0) at (0,1) {A-IoT Reader};
  \node (d0) at (6,1) {A-IoT Device};

  % Lifelines
  \draw[dashed] (0,0) -- (0,-5.6);
  \draw[dashed] (6,0) -- (6,-5.6);

  % 1) Paging (CBRA)
  \draw[->] (0,0) -- (6,0)
    node[midway, above, msg]
    {A-IoT Paging\\
     \footnotesize AccessType = CBRA, TransactionID = $\tau$, [PagingID]};
  \node[noteR] at (6.15,-0.4) {%
    Device checks PagingID and stores $\tau$ as the new Transaction ID
    (if the old procedure failed or $\tau$ is new).};

  % 2) Access Random ID (Msg1)
  \draw[->] (6,-1.5) -- (0,-1.5)
    node[midway, above, msg]
    {Access Random ID (MSG1)\\
     \footnotesize RandomID = $r$ (16-bit),\\
     \footnotesize selected access occasion};
  \node[noteL] at (-0.15,-1.5) {%
    Reader receives RandomID $r$ from one or more devices.};

  % 3) Random ID Response (Msg2)
  \draw[->] (0,-2.8) -- (6,-2.8)
    node[midway, above, msg]
    {Random ID Response (MSG2)\\
     \footnotesize \{Echoed RandomID = $r$, [Assigned ID]\},\\
     \footnotesize TransactionID = $T$};
  \node[noteR] at (6.15,-2.8) {%
    Device finds the entry matching its $r$ (and applies any frequency shift if present).
    If Assigned ID is present: ID $\leftarrow$ Assigned ID; else: ID $\leftarrow r$.};

  % 4) D2R upper-layer data
  \draw[->] (6,-4.0) -- (0,-4.0)
    node[midway, above, msg]
    {D2R Upper Layer Data Transfer\\
     \footnotesize (ID, data segment, MoreDataIndicator = 0/1)};
  \node[noteL] at (-0.15,-4.0) {%
    Reader receives data associated with ID under Transaction $\tau$.};

  % 5) Optional feedback / scheduling
  \draw[->] (0,-5.2) -- (6,-5.2)
    node[midway, above, msg]
    {R2D Upper Layer Data Transfer / NACK\\ 
    Feedback \footnotesize (status, next D2R scheduling info)};
  \node[noteR] at (6.15,-5.3) {%
    Reader acknowledges or requests retransmission, and may provide updated D2R scheduling information.};

\end{tikzpicture}
\end{adjustbox}
\caption{Example CBRA-based A-IoT MAC procedure: paging, random access, ID assignment, and D2R data transfer.}
\label{fig:cbra_aiot_proc}
\end{figure*}
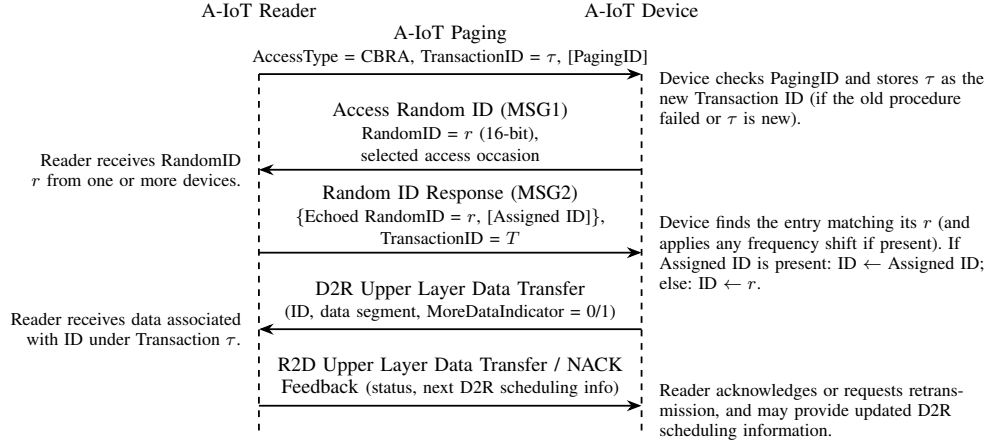

Before a device can transmit its report, it must successfully complete an \gls{aiot} \gls{cbra} handshake, which is initiated by an \gls{r2d} paging message from the reader in each paging round~\cite{3gppTS38391}, as shown in \figurename~\ref{fig:cbra_aiot_proc}. Specifically, the paging message targets connected devices and configures a set of candidate \glspl{ao} for the subsequent \gls{d2r} transmission of the Access Random ID (MSG1). Each paging round offers $R$ orthogonal \gls{cbra} resources (\glspl{ao}), which can be interpreted as time-frequency opportunities reserved for \gls{ra}.

\Glspl{ao} are not referenced to a globally fixed-slot structure such as \gls{tti}. Instead, the beginning of each \gls{ao} set is explicitly indicated by an \gls{r2d} paging message, which also indicates whether additional \glspl{ao} will follow in subsequent trigger messages. 
Each triggered \gls{aiot} device then performs a \emph{one-shot} random selection of an \gls{ao} within this configured set by drawing a discrete uniform index $i\in\{0,\ldots,R-1\}$. Note that the index $i$ 
%is \emph{not} updated across triggers and 
does not represent a contention back-off. Instead, the device uses a counter only as an implementation aid to \emph{locate} the preselected \gls{ao} within the same paging round in an asynchronous, trigger-defined sequence. When the counter reaches zero, the device transmits MSG1 in that \gls{ao} using a 16-bit random ID. After MSG1 is transmitted or the access opportunity expires, the counter is reset and a new index $i$ is drawn the next time the device is triggered. 
%Hence, the device identifies the timing of its selected \gls{ao} by monitoring this trigger sequence and counting \glspl{ao} within the current paging round.

We assume that a device can attempt \gls{cbra} only if it has sufficient stored energy at the paging instant as indicated earlier. 
%In particular, a device is \emph{eligible} to participate in \gls{cbra} in a given paging round if its stored energy satisfies $E_\tau \geq E_{\min}$ (equivalently, if a voltage-domain model is used, $V_\tau \geq V_{\min}$), as shown in \figurename~\ref{fig:energy_buffer}. 
Such \textit{eligible} devices spend $E_{\mathrm{RA}}$ energy units attempting \gls{cbra}. The reader listens to all $R$ \glspl{ao}, and if exactly one device selects a particular \gls{ao}, the corresponding access attempt is successful, and the device may proceed with the remaining handshake and report delivery. If two or more devices select the same \gls{ao}, a collision occurs, and we assume that all attempts mapped to that \gls{ao} fail. In that case, a device may re-attempt in subsequent paging rounds, subject to its energy availability. 

Each trigger-defined set provides 
%$R \triangleq X  F$ 
$R$ \glspl{ao}, 
%where $X$ is the number of time-domain resources and $F$ is the number of frequency-domain resources. 
defined by the number of time/frequency-domain resources.
If a paging round contains $Y$ trigger-defined sets, the total number of \glspl{ao} configured in that paging round is $Y  R$. For analytical convenience, we index paging rounds by a sequence number $\tau = 1,\ldots, T$, being $T$ the total number of paging rounds during the simulation. 
% Thus, the total number of \glspl{ao} per simulation is gT. 
In each paging round, every energy-eligible device in the polled group attempts the paging-triggered \gls{cbra} round with an 
%fixed 
access probability %$q=\min\!\left(1, {R/N}\right)$ 
$q$ and, conditioned on attempting, selects one of the configured \glspl{ao} uniformly with probability $1/R$. This baseline captures a classical slotted-ALOHA access rule. 
An access attempt succeeds only if exactly one device selects the corresponding \gls{ao}; otherwise, a collision occurs and all devices mapped to that \gls{ao} fail and may retry in a later paging round, subject to energy availability.

\subsection{Energy consumption and harvesting model}\label{sec:capacitor}

We assume harvested energy is stored in a capacitor with capacitance $C$. The capacitor charges when the harvesting current $I_h$ exceeds the load current $I_c$, and discharges when the load draws more current than is harvested. The load includes all energy-consuming elements of the system, including the MCU, sensors, radio, and inherent capacitor leakage. They are collectively modeled as an equivalent resistance $R_{eq}$, which for a device state $s$ is calculated as 
\begin{align}
    R_{eq}^{(s)} = \frac{V_{supply}}{I_c^{(s)}},
\end{align}
where $V_{supply}$ represents the constant supply voltage, and $I_c^{(s)}$ the current consumption in state $s$. If the device spends $\Delta t$ time in state $s$, starting from voltage $V_0$, then the voltage at the end of the interval is calculated as
\begin{align}
    \label{eq:voltage_change}
    V^{(s)}(\Delta t, V_0) = I_h  R_{eq}^{(s)}  \left(1-e^{-\frac{\Delta t}{ R_{eq}^{(s)}  C}}\right) + V_0  e^{-\frac{\Delta t}{ R_{eq}^{(s)}  C}},
\end{align}
where, without loss of generality, $I_h$ is considered constant during the time interval $\Delta t$. 
%\textcolor{red}
{To model $I_h$, we use \gls{eh} data from the power management integrated circuit collected in~\cite{nasser2025feasibility}.}   
Finally, the energy $E$ stored in the capacitor is directly related to its voltage $V$ as follows
\begin{align}
    E = \frac{1}{2}  C  V^2.
\end{align}

%----------------------------------------------------------------------

\subsection{Operational states and transition probabilities}

An \gls{aiot} device $i$ can operate in one of three states: (i) \emph{``OFF''}, in which the device is not available for paging or \gls{ra}, as $V^{(i)} < V_{\min}$; (ii) \emph{``idle''}, wherein the device is monitoring for paging occasions; and (iii) \emph{``RA''}, wherein it is contending for uplink access using \gls{ra}. \gls{aiot} devices can harvest energy whenever it is available, regardless of their operational state. 
Herein, we adopt a reduced two-state model ${{S^{(i)}\in {\mathcal{S}}}=\{S_1, S_2\}}$, where $S_1$ denotes the {\em``idle'' state} and $S_2$ the {\em``RA'' state}. The effect of the \emph{``OFF''} mode is represented into the energy-availability probability $p_e^{(i)}=\Pr(V_\tau^{(i)} \ge V_{\min}),$ which captures whether device $i$ is eligible to react to the paging message at round $\tau$. Hence, the two-state Markov chain is conditioned on the device being energy-available at the paging instant.  

We denote $p_{k,l}^{(i)}$ as the transition probability for a device $i$ from a state $S_k$ into a state $S_l$. The states and their transition probabilities are described through a discrete-time Markov process per paging round, as shown in Fig.~\ref{fig:system_model}. 
The state evolution follows the first-order Markov property:
\begin{align}
\Pr(X_{n+1} = S_\ell \mid X_n = S_k, X_{n-1}, \dots)
= \nonumber\\ 
\Pr(X_{n+1} = S_\ell \mid X_n = S_k)
= p^{(i)}_{k,\ell},
\end{align} 
where $p^{(i)}_{k,\ell}$ is the transition probability of device $i$, and $k,\ell \in {1,2}$. The transition probability matrix for a device $i$ is
\begin{equation}
\mathbf{P}^{(i)} =
\begin{bmatrix}
p^{(i)}_{1,1} & p^{(i)}_{1,2} \\
p^{(i)}_{2,1} & p^{(i)}_{2,2}
\end{bmatrix},
\quad
\sum_{\ell=1}^{2} p^{(i)}_{k,\ell} = 1, \ \forall k.
\end{equation}
Note that $p_{2,1}^{(i)} = 1$, $p_{2,2}^{(i)} = 0$ and $p_{1, 1}^{(i)} = 1 - p_{1,2}^{(i)}$.

%----------------------------------------------------------------------

\section{Adaptive \gls{cbra} \& \gls{eh}-Aware Control}\label{sec:ra}

%We consider three access policies. 
The access policy determines how devices contend for \glspl{ao} and how the reader regulates the access in each paging round.
% \subsection{Ideal link} 
% An ideal scheduler assigns distinct \glspl{ao} to all contending devices, resulting in collision-free access. If the number of competing devices exceeds the number of available \glspl{ao}, the scheduler randomly selects a number of devices matching the available \glspl{ao} and assigns one to each to prevent collisions during that paging round. While this approach is not feasible in practice, it serves as an upper bound for performance.
% \subsection{Uncontrolled \gls{cbra}} 
% Every energy-eligible device in the polled group attempts the paging-triggered \gls{cbra} round with a fixed access probability $q=\min\!\left(1, {R/N}\right)$ and, conditioned on attempting, selects one of the configured \glspl{ao} uniformly with probability $1/R$. This baseline captures a classical slotted-ALOHA access rule. 
% An access attempt succeeds only if exactly one device selects the corresponding \gls{ao}; otherwise, a collision occurs and all devices mapped to that \gls{ao} fail and may retry in a later paging round, subject to energy availability.
\begin{algorithm}[t]
\caption{\gls{eh}-aware access probability control}
\label{alg:cbra_adaptive}
\scriptsize
\small
\begin{algorithmic}[1]
\State \textbf{Input:} $R$, $Y$, ${K}$, $V_{\min}$, \gls{eh} profile  
\State \textbf{Maintain:} predicted voltages $\{\widehat{V}_\tau^{(i)}\}_{i=1}^N$
\For{each paging round $\tau=1,\ldots,T$}
    \State Update $\widehat{V}_\tau^{(i)}$ for all $i$ using \eqref{eq:voltage_change} 
    \State Compute ${\widehat{K p_e}(\tau)}$ and Set $q_\tau \leftarrow \min\!\left(1, \frac{R}{\widehat{K p_e}(\tau)}\right)$
    \State Broadcast $q_\tau$ in the \gls{r2d} paging message
    \State Each polled device $i$ with $V_\tau^{(i)}\ge V_{\min}$, attempt \gls{cbra} 
    \Statex {\hspace{5mm}with probability $q_\tau$}
    %\State Observe number of decoded MSG1/MSG2 outcomes and 
    %\Statex {\hspace{5mm}refine $\eta$ and $\widehat{V}_{\tau+1}^{(i)}$}
\EndFor
\end{algorithmic}
\end{algorithm}
%\subsection{Adaptive \gls{cbra} control}\label{sec:proposal}
In this paper, we propose an \gls{eh}-aware access control in which the reader broadcasts an access probability $q_\tau$ in each paging message. Upon reception, each energy-available paged device decides whether to attempt \gls{cbra} according to $q_\tau$. The reader computes this probability from historical access observations and predicted device energy levels to regulate the offered load per \gls{ao} and reduce collisions.
We introduce a lightweight reader-side mechanism for adaptive \gls{cbra} control. The main idea is to regulate the number of devices contending in the paging round $\tau$ by adjusting the access probability $q_\tau$. To this end, the reader estimates the expected number of energy-available devices in the targeted paging group, denoted by $\widehat{K p_e}(\tau)$, using the capacitor model in Section~\ref{sec:capacitor}, the \gls{eh} profile, and the device activity history. Devices report their voltage $V_\tau^{(i)}$ and message attempt numbers during each transmission. The receiver obtains these parameters when the transmission is successful, using them to improve future predictions.

The control objective is to keep the expected number of contenders per \gls{ao} close to one, thereby maximizing access efficiency while limiting collisions. Let $K'$ denote the number of devices that actually participate in a paging round. Assuming uniform \gls{ao} selection, the expected number of successful contenders per round is
\begin{align}\label{eq:P_succ}
\mathbb{E}[N_{\mathrm{s}}] = K'\left(1-\frac{1}{R}\right)^{K'-1}
\approx K' e^{-K'/R}, 
\end{align}
%where the approximation holds for sufficiently large $R$, and is maximized when $K' = R$. 
which can be maximized when
$K'\approx R$.
Accordingly, the access probability is selected as
\begin{align}
q_\tau = \min\!\left(1, \frac{R}{\widehat{K p_e}(\tau)}\right),
\label{eq:q_rule}
\end{align}
where $\widehat{K p_e}(\tau)$ is the reader's estimate of the number of energy-available devices in the addressed group. Algorithm~\ref{alg:cbra_adaptive} summarizes the proposed procedure. 
The reader estimates how many of the paged devices are likely to have sufficient energy at the beginning of paging round $\tau$. Specifically, the reader uses the most recent report and the time elapsed since the last update to compute $\widehat{V}_\tau^{(i)}$. A device is considered available if its predicted voltage satisfies $\widehat{V}_\tau^{(i)} \geq V_{\min}$. Therefore, the estimated number of energy-available devices within the paging group is 
\begin{align}
\widehat{Kp_e}(\tau)=\sum_{i\in\mathcal{K}_\tau}\mathbf{1}\{\widehat{V}_\tau^{(i)}\geq V_{\min}\}.
\label{eq:q_rule2}
\end{align}

%-----------------------------------------------------------------

\section{Results Analysis}\label{sec:results}
\begin{table}[t]
\centering
\caption{Simulation parameters}
\label{tab:simulation}
\begin{tabular}{l c l}
\hline
\textbf{Parameter} & \textbf{Symbol} & \textbf{Value} \\
\hline
Capacitance & $C$ & $[1,10]$ F\\
\gls{cbra} \glspl{ao} per trigger-defined set & $R$ & $\{8,16,32\}$ \\
Leakage current & $I_{\ell}$ & $0.03~\text{mA}$ \\
MCU active current (during sensing/TX) & $I_{\text{mcu}}$ & $0.091~\text{mA}$ \\
Minimum voltage threshold & $V_{\min}$ & $1.8~\text{V}$ \\
%Paging rounds per run & $T$ & $ $ \\
Polling probability & $p_{r2d}$ & $0.01$ \\
Sensor I\textsuperscript{2}C time & $T_{\text{sht,i2c}}$ & $0.000325~\text{s}$ \\
Sensor measurement time & $T_{\text{sht,meas}}$ & $0.0055~\text{s}$ \\
Simulation/paging step & $T_i$ & $5~\text{s}$ \\
Sleep current (including  \gls{wur})    &$I_{\text{sleep}}$     &$0.02$ mA\\
Supply voltage & $V_{\text{supply}}$ & $3.3~\text{V}$ \\
Transmit current (radio) & $I_{\text{tx}}$ & $20.65~\text{mA}$ \\
Trigger-defined sets per paging round & $Y$ & $1$ \\
\hline
\end{tabular}
\end{table}

\begin{figure*}[t]
\centering
% {\includegraphics[width=0.325\linewidth]{fig/P_collision_R_8.png}}
% {\includegraphics[width=0.325\linewidth]{fig/P_collision_R_16.png}}
% {\includegraphics[width=0.325\linewidth]{fig/P_collision_R_32.png}}
% {\includegraphics[width=0.325\linewidth]{fig/msg2_per_RA_R_8.png}}
% {\includegraphics[width=0.325\linewidth]{fig/msg2_per_RA_R_16.png}}
% {\includegraphics[width=0.325\linewidth]{fig/msg2_per_RA_R_32.png}}
    \begin{subfigure}[b]{0.32\linewidth}
        \centering
        \includegraphics[width=\linewidth]{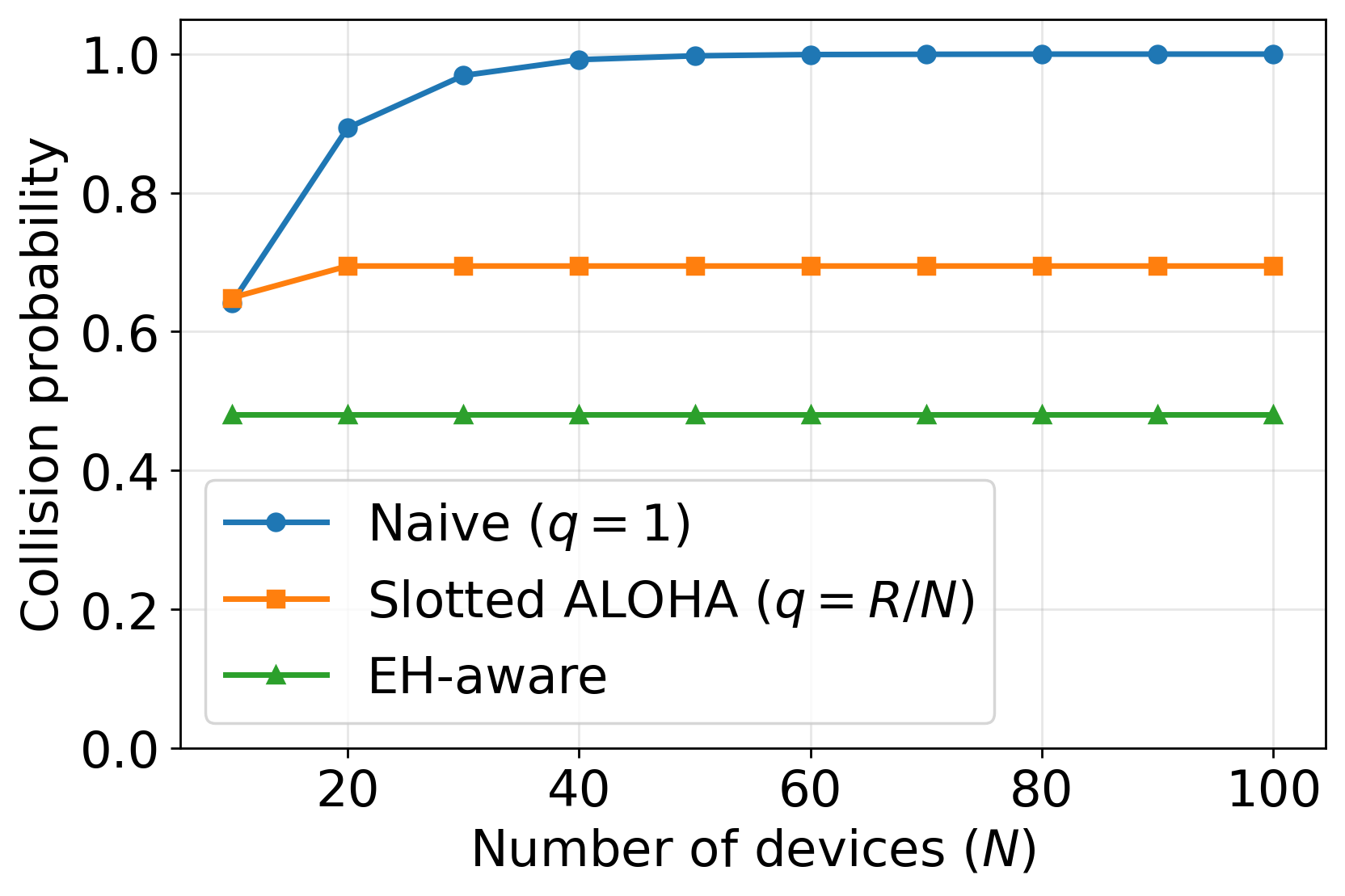}
    \end{subfigure}
    \begin{subfigure}[b]{0.32\linewidth}
        \centering
        \includegraphics[width=\linewidth]{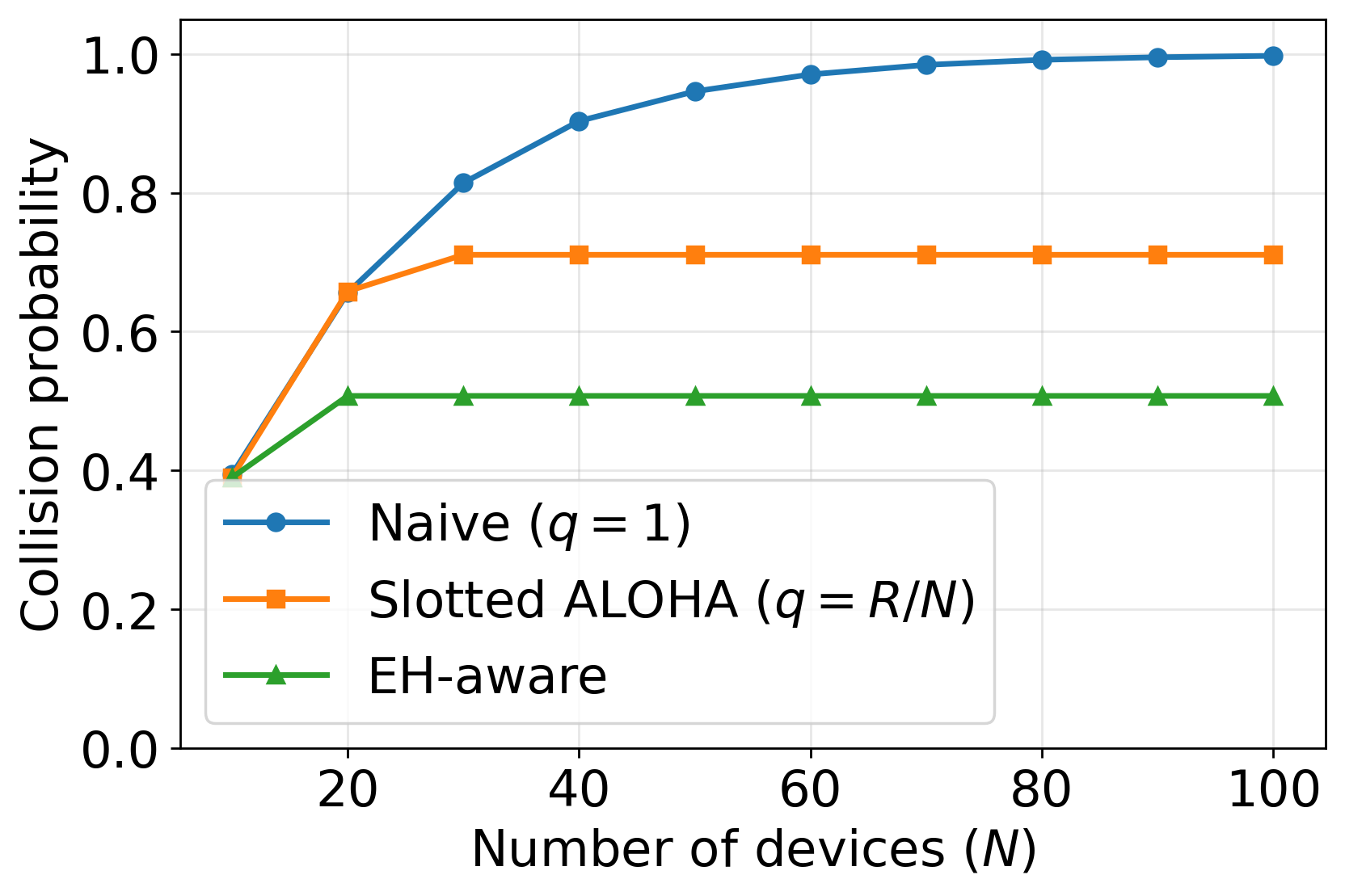}
    \end{subfigure}
    \begin{subfigure}[b]{0.32\linewidth}
        \centering
        \includegraphics[width=\linewidth]{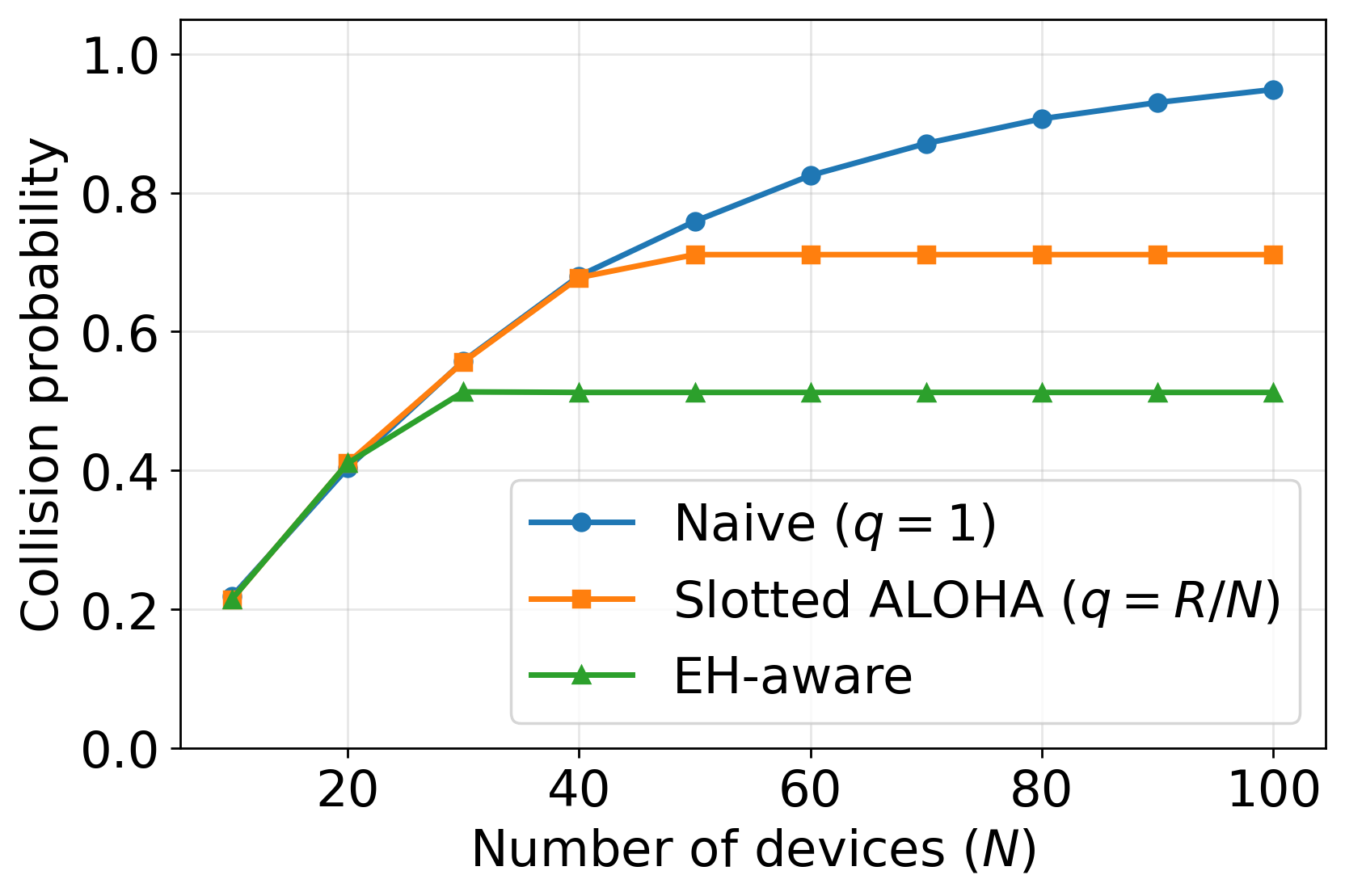}
    \end{subfigure}
    %\vspace{1mm}
    \begin{subfigure}[b]{0.32\linewidth}
        \centering
        \includegraphics[width=\linewidth]{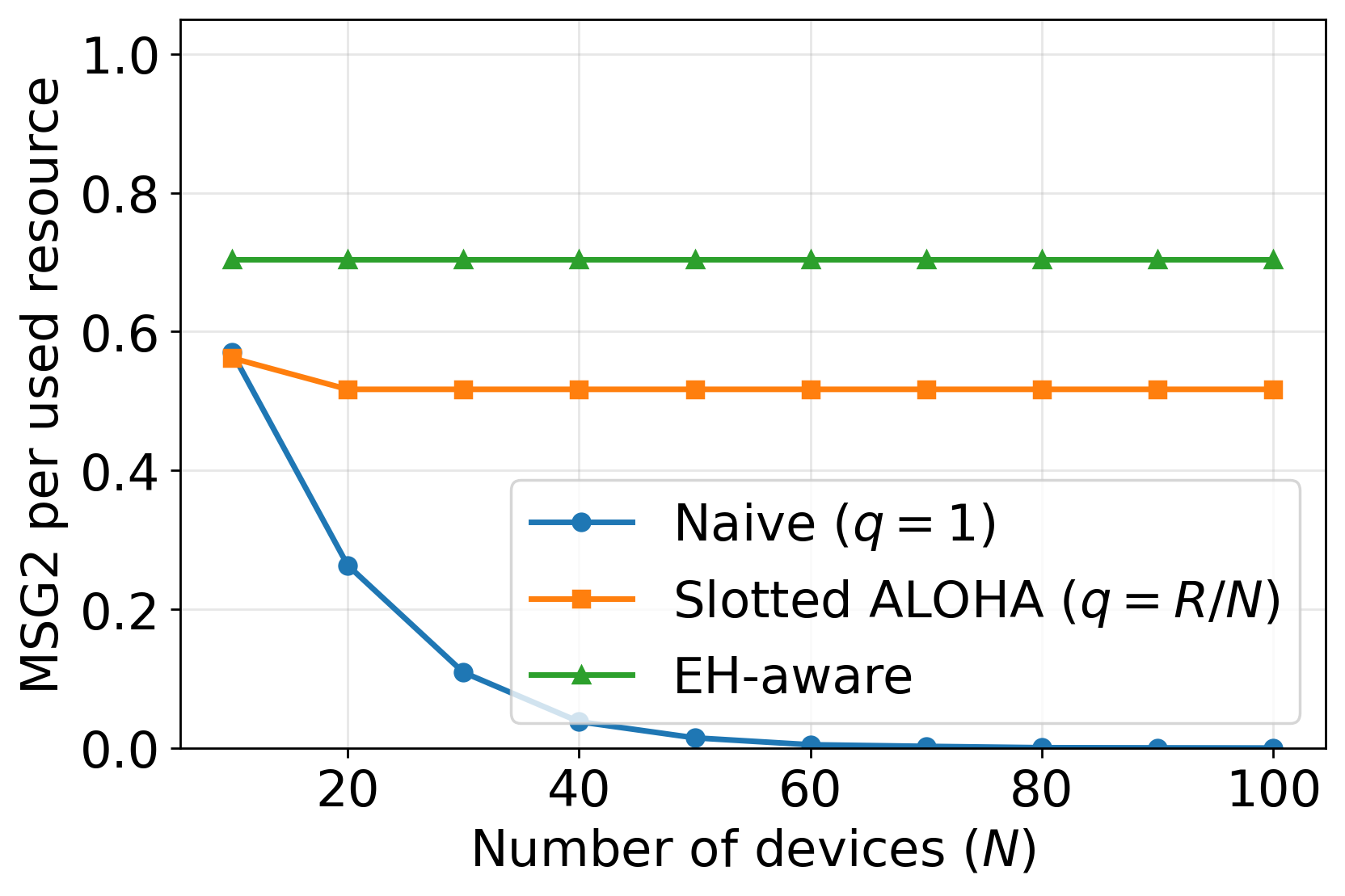}
        \caption{$R=8$}
    \end{subfigure}
    \begin{subfigure}[b]{0.32\linewidth}
        \centering
        \includegraphics[width=\linewidth]{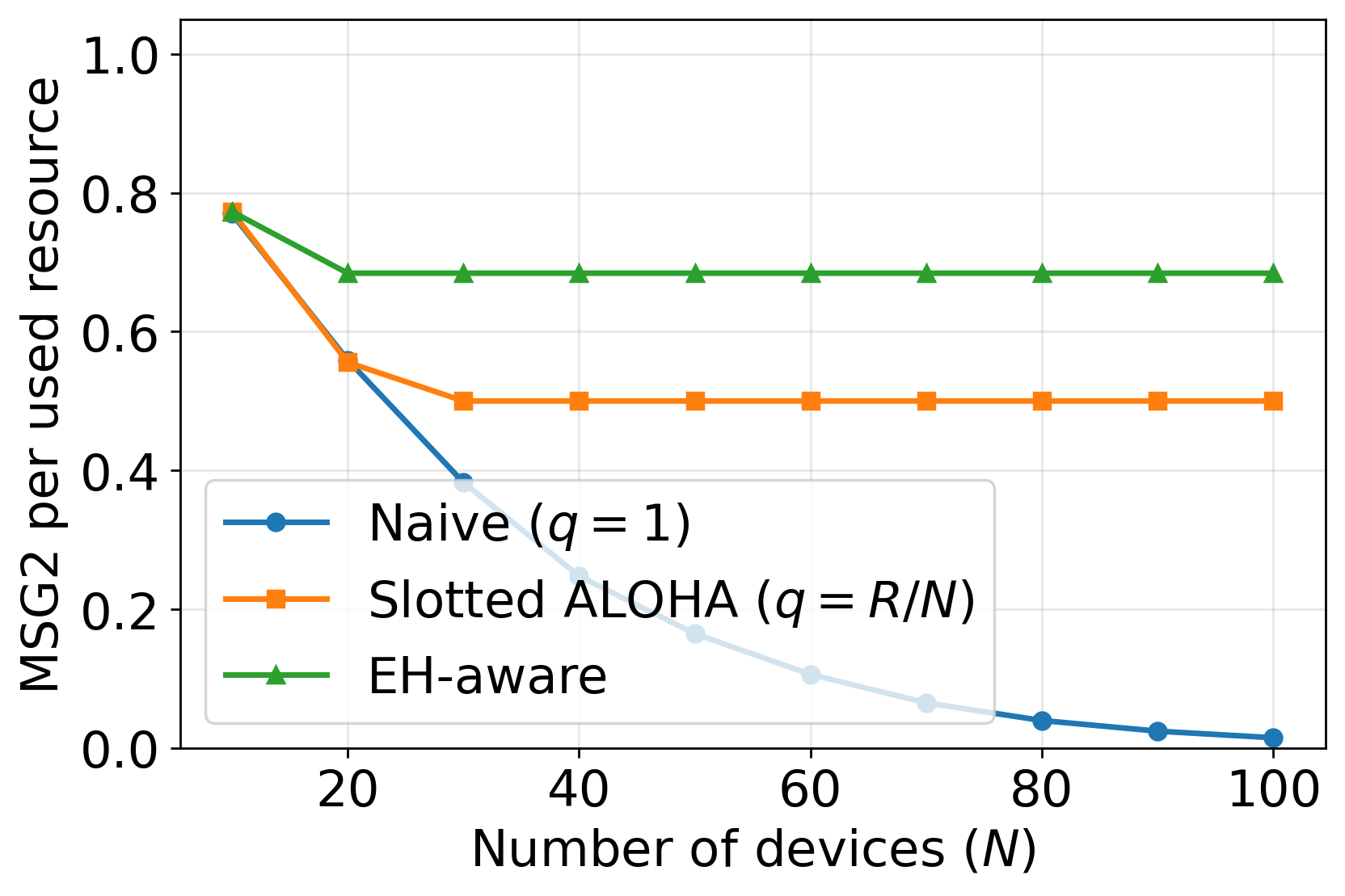}
        \caption{$R=16$}
    \end{subfigure}
    \begin{subfigure}[b]{0.32\linewidth}
        \centering
        \includegraphics[width=\linewidth]{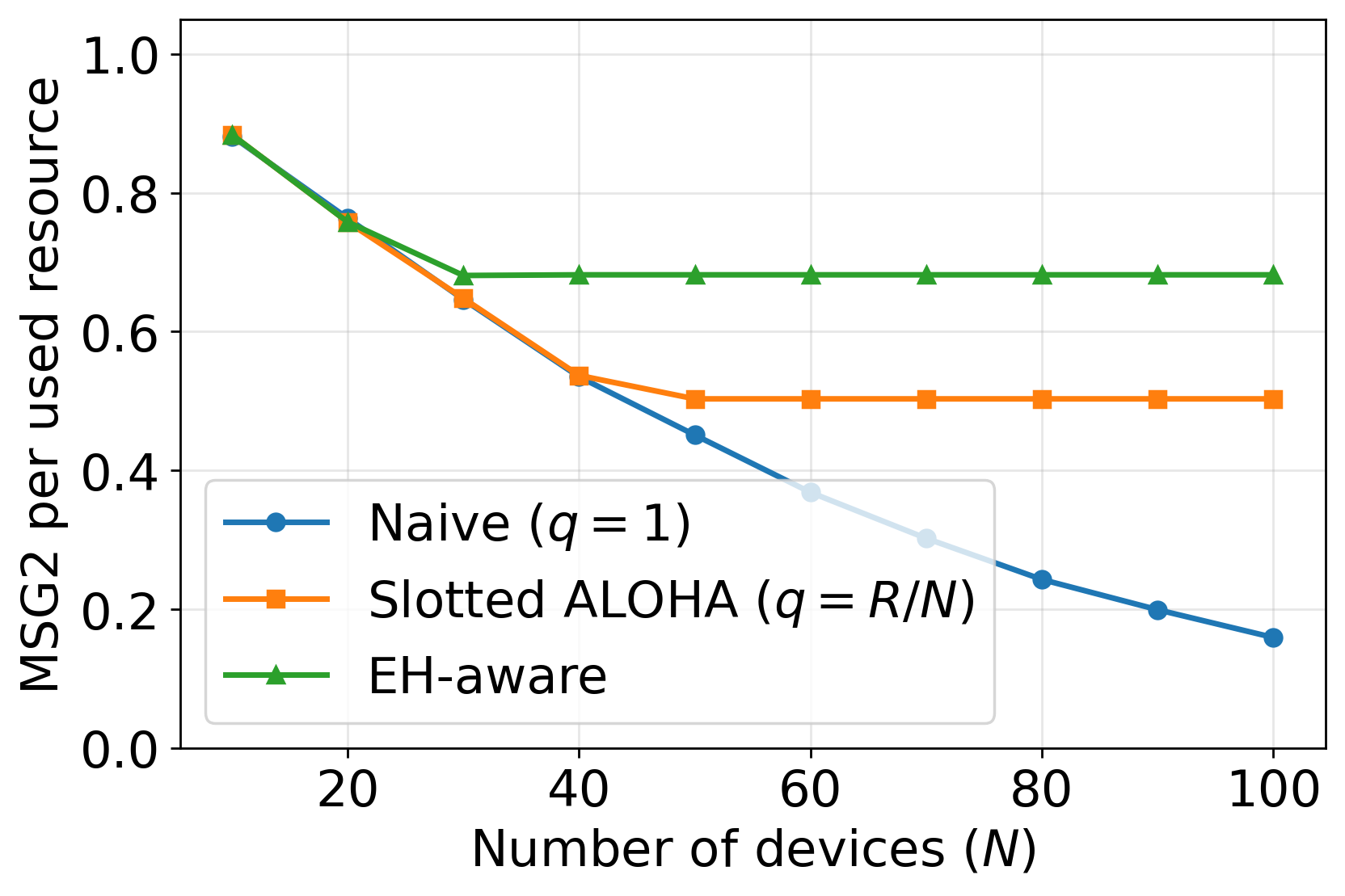}
        \caption{$R=32$}
    \end{subfigure}
\caption{(Top) Probability of collision and (bottom) mean number of successful MSG2 receptions per used \glspl{ao} as a function of the number of devices ($N$) for $R \in \{8,16,32\}$.}
\label{fig:Pcol_msg2}
\vspace{-3mm}
\end{figure*}

We consider an \gls{aiot} deployment with ${N} \in [10, 100]$. We perform $250$ runs of Monte Carlo simulations, each one lasting $2\times 10^5$ simulation steps. Moreover, we assume $Y=1$, $p_{\mathrm{r2d}}=0.01$, and interval steps of 5 s, \textit{i.e.}, on average one paging every 100 rounds (500 s). Table~\ref{tab:simulation} summarizes the parameters used in simulations assuming a SHT30 temperature and humidity sensor and a STM32L496 \gls{mcu}~\cite{nasser2025feasibility,stmicroelectronics2024stm32l496xx} unless explicitly stated otherwise. 

%subsection{Performance metrics}

The per-round probability that a polled device successfully completes the \gls{cbra} MSG1 transmission depends on both energy availability and contention. Since an energy-available device attempts access with probability $q_\tau$, a device $i$ attempts \gls{cbra} with probability  
\begin{equation}
p_{1,2}^{(i)}(\tau) = p_e^{(i)}  q_\tau.
\end{equation}
This expression captures the joint effect of intermittent energy and reader-side access control. In particular, increasing $q_\tau$ raises the number of contenders and thus the collision probability, whereas choosing $q_\tau$ too conservatively may under-utilize the available \glspl{ao}. 

An ideal scheduler assigns distinct \glspl{ao} to all contending devices, resulting in collision-free access. If the number of competing devices exceeds the number of available \glspl{ao}, the scheduler randomly selects a number of devices matching the available \glspl{ao} and assigns one to each to prevent collisions during that paging round. While this approach is not feasible in practice, it serves as an upper bound for performance.

\subsection{Performance metrics}

For each policy, we evaluate three performance indicators: 
\textbf{(i)} the reporting collision probability per \gls{cbra} attempt, $P_{\mathrm{col}}=1-(1-1/R)^{K'-1}$; 
\textbf{(ii)} the mean number of successful contenders receptions per used resource in each \gls{ao} ($Y  R$),  defined in \eqref{eq:P_succ}; and 
\textbf{(iii)} the mean number of paging rounds required to report a successful MGS2 from an arbitrarily selected device, $N/\mathbb{E}[N_{\mathrm{s}}]$.
%\begin{itemize}
    %\item[i)] The reporting collision probability per \gls{cbra} attempt, $P_{\mathrm{col}}=1-(1-1/R)^{K'-1}$.
    % \item[ii)] The mean number of successful contenders receptions per used resource in each \gls{ao} ($Y  R$),  defined in \eqref{eq:P_succ}.
    % \item[iii)] The mean number of paging rounds required to report a successful MGS2 from an arbitrarily selected device, $N/\mathbb{E}[N_{\mathrm{s}}]$.
%\end{itemize}    
These metrics quantify the trade-offs between reliability and access efficiency in \gls{aiot} networks using \gls{cbra}.
Note that the collision-probability and resource-efficiency metrics in \figurename~\ref{fig:Pcol_msg2} are shown only for the three non-ideal schemes, a naive baseline with $q=1$, a static slotted-ALOHA baseline with $q=R/N$, and the proposed EH-aware policy, since the ideal benchmark yields zero collisions and unit MSG2-per-used-\gls{ao} efficiency. The ideal benchmark is therefore included only in the delay-oriented metric of \figurename~\ref{fig:aveRA}, where the number of paging rounds still depends on the number of contenders and the finite \gls{ao} budget.

\subsection{Simulation results}

\figurename~\ref{fig:Pcol_msg2} reports the collision probability (top row) and the mean number of successful MSG2 receptions per used \gls{ao} (bottom row). The naive baseline 
%($q_\tau=1$) 
becomes rapidly collision-limited as the number of devices increases. For $R=8$, its collision probability grows from about $0.65$ at low load to nearly one for large $N$, while for $R=16$ and $R=32$ the same trend remains visible, although shifted to higher device counts due to the larger \gls{ao} budget. In contrast, the static slotted-ALOHA baseline ($q=R/N$) keeps the collision probability significantly below the naive scheme and prevents the near-saturation regime, but still operates at a comparatively high and nearly constant collision level. Notably, the proposed \gls{eh}-aware control achieves the lowest collision probability and regulates the offered load per \gls{ao} while maintaining a nearly flat collision level over the full range of $N$. This stabilization effect becomes increasingly relevant as $N$ grows, since the gap between \gls{eh}-aware and naive operation widens for all considered $R$ values. 
The bottom row of \figurename~\ref{fig:Pcol_msg2} illustrates a similar trend from a resource-efficiency perspective. Under the \gls{eh}-aware policy, the mean number of successful MSG2 receptions per used \gls{ao} remains comparatively stable as $N$ increases. In contrast, the naive scheme experiences a significant efficiency decline because a larger proportion of resources is consumed by collided attempts. Note that this degradation is more pronounced with smaller \gls{ao} budgets and becomes less severe as $R$ increases. While the static slotted ALOHA baseline shows a marked improvement compared to the naive scheme, it still exhibits lower efficiency when compared to the EH-aware control proposal.  
Nevertheless, even for $R=32$, the \gls{eh}-aware policy preserves a clear efficiency advantage at medium and high device densities.

\begin{figure}[t]
\centering
{\includegraphics[width=0.75\linewidth]{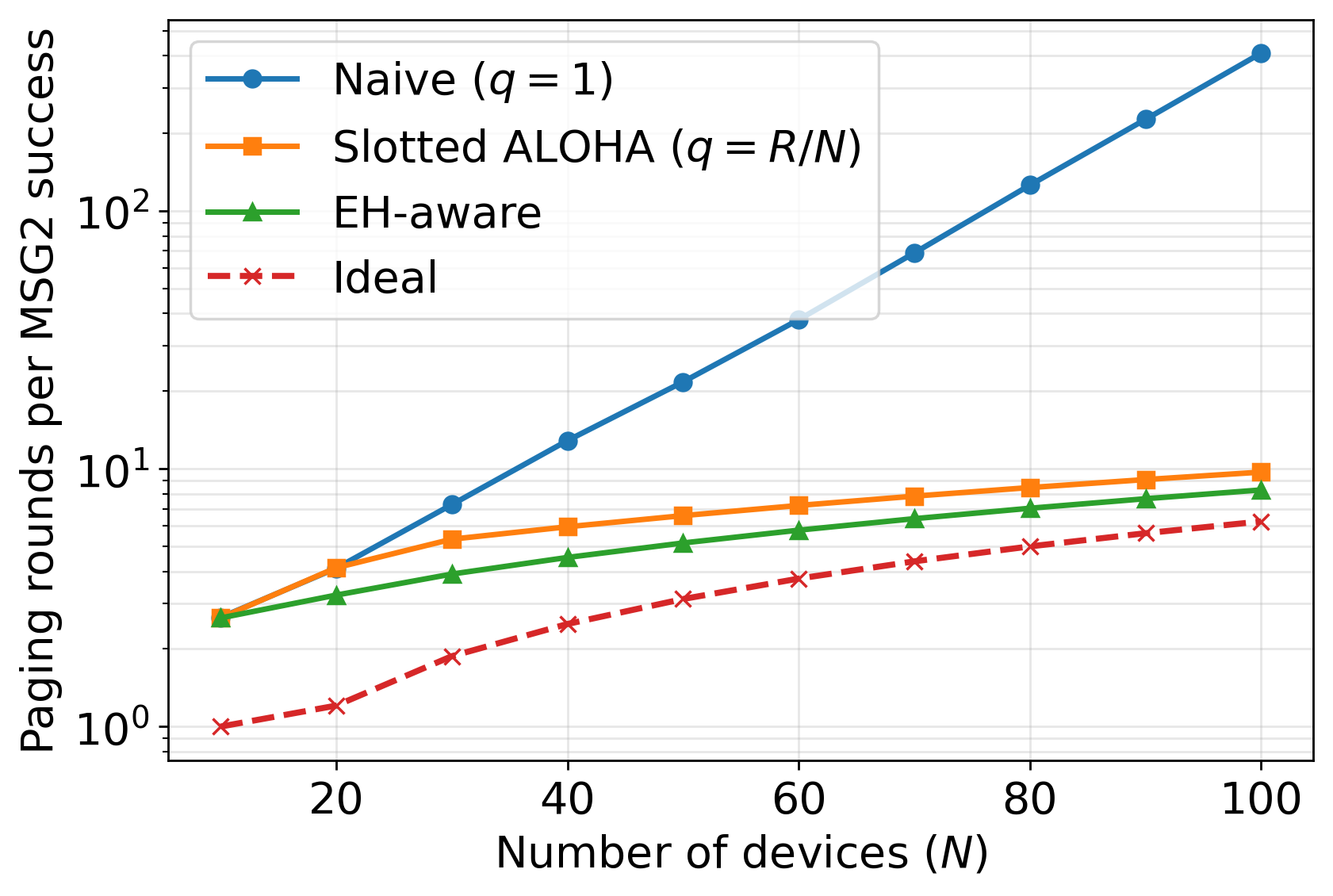}}
{\includegraphics[width=0.75\linewidth]{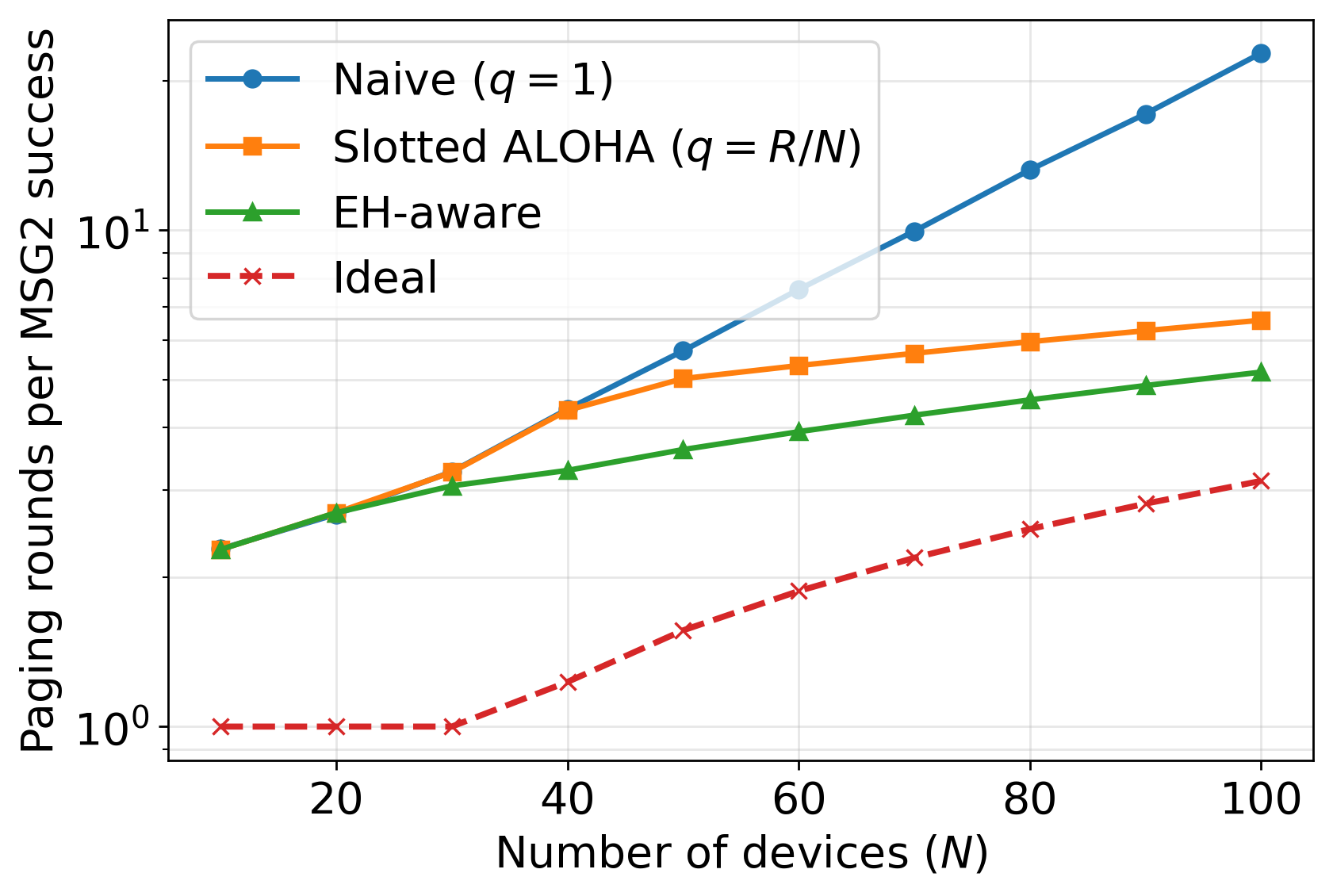}}
\caption{Mean number of paging rounds ($\tau$) to receive a successful MSG2 from each device as a function of the number of devices ($N$) for $R=16$ (top) and $R=32$ (bottom).}
\label{fig:aveRA}
\vspace{-3mm}
\end{figure}
\figurename~\ref{fig:aveRA} evaluates the mean number of paging rounds required to receive one successful MSG2 per device. Unlike the two metrics in \figurename~\ref{fig:Pcol_msg2}, this quantity is also shown for the ideal benchmark, because even in the absence of collisions, the delay still depends on the number of contenders and the finite number of available \glspl{ao} per paging round. The \gls{eh}-aware policy remains significantly closer to this ideal behavior than the naive and slotted ALOHA schemes. For both $R=16$ and $R=32$, the naive baseline exhibits exponential growth in required paging rounds as $N$ increases, indicating repeated collision-driven re-access attempts. In contrast, the \gls{eh}-aware and the slotted ALOHA policies grow much more gradually with $N$. Once the offered \glspl{ao} start approaching one contender per paging round, the naive scheme quickly accumulates repeated collisions, which translates into a steeper increase in the average number of paging rounds per successful MSG2. This trend illustrates that regulating the number of participating devices is effective not only in reducing collisions, but also in reducing access delay and repeated \gls{cbra} attempts.
The results show a clear trade-off between \gls{ao} budget and access control. Increasing $R$ alleviates contention for both policies; however, the proposed \gls{eh}-aware control prevents overloading the random-access pool when the number of polled and energy-available devices increases. 

\section{Conclusion}\label{sec:conclusions}
In this paper, we examined asynchronous \gls{cbra} uplink reporting in \gls{aiot} networks using a Bernoulli-based reader polling. We evaluated four access policies: an ideal collision-free benchmark, a naive \gls{cbra} baseline in which every energy-available device attempts access ($q=1$), a static slotted-ALOHA baseline with $q=R/N$, and a proposed EH-aware control in which the reader broadcasts an access probability to regulate the offered load per \gls{ao}. The results indicated that the naive policy becomes significantly limited by collisions as the number of devices increases, which greatly reduces both resource efficiency and access delay. 
The static slotted-ALOHA baseline alleviates this effect by limiting the number of contenders per \gls{ao}, but it still operates with a relatively high collision level.
In contrast, the proposed \gls{eh}-aware method maintains a nearly constant level of collisions and achieves a considerably higher number of successful MSG2 receptions per used \gls{ao}, while also requiring substantially fewer paging rounds for each successful MSG2 compared to the two fixed baselines. These findings suggest that lightweight access-probability control at the reader side is an effective mechanism for enhancing the reliability and efficiency of \gls{aiot} reporting under conditions of intermittent energy availability. 

% conference papers do not normally have an appendix

% use section* for acknowledgment
\section*{Acknowledgment}
This work has been partially supported by the Research Council of Finland (Grants 369116 (6G Flagship) and 362782 (ECO-LITE)), the Finnish Foundation for Technology Promotion, and the European Commission through the Horizon Europe/JU SNS project AMBIENT-6G (Grant 101192113).

\bibliographystyle{IEEEtran}
\bibliography{bib}

% \clearpage
% \input{Simulation_values}

% \clearpage
% \input{CFA}

\end{document}